\documentclass{aastex7}

\newcommand{\tess}{{\it TESS}}

\newcommand{\gaia}{{\it Gaia}}
\newcommand{\grp}{$G_{\text{RP}}$}

\newcommand{\kms}{km\,s$^{-1}$}

\newcommand{\mjup}{\mbox{M$_{J}$}}
\newcommand{\rjup}{\mbox{R$_{J}$}}

\newcommand{\msun}{\mbox{M$_{\odot}$}}
\newcommand{\rsun}{\mbox{R$_{\odot}$}}

\newcommand{\teff}{$T_{\rm eff}$}
\newcommand{\feh}{\mbox{[Fe/H]}}

\newcommand{\logg}{$\log g$}
\newcommand{\vsys}{$V_{\text{sys}}$}

\newcommand{\phngts}{PHNGTS}
\newcommand{\mysim}{\mathord{\sim}}

\newcommand{\allesfitter}{\texttt{allesfitter}}

\newcommand{\dragons}{\texttt{DRAGONS}}

\newcommand{\ispec}{\texttt{iSpec}}
\newcommand{\mad}{\textit{MAD}}

\newcommand{\chisq}{$\chi^2$}
\newcommand{\redchisq}{$\chi^2_{\nu}$}

\newcommand{\nigelid}{TIC-165227846}
\newcommand{\nigelgid}{Gaia\,DR3\,3459141135209729536}
\newcommand{\nigelsysname}{NGTS-EB-8}
\newcommand{\nigelAname}{NGTS-EB-8\,A}
\newcommand{\nigelBname}{NGTS-EB-8\,B}

\defcitealias{Bryant2023OccurrenceGiantsMDwarfs}{B23}
\defcitealias{Baraffe2015Isochrones}{B15}
\defcitealias{Schweitzer2019CARMENESTargetStars}{S19}
\defcitealias{Mann2015ConstrainMDwarfs}{M15}
\defcitealias{Mann2019ConstrainMdwarfII}{M19}

\newcommand{\Brrfull}{$0.574^{+0.038}_{-0.042}$}
\newcommand{\Brsumafull}{$0.0556[7]$}
\newcommand{\Bcosifull}{$0.032^{+0.001}_{-0.001}$}
\newcommand{\Bepochfull}{$2460054.77781[6]$}
\newcommand{\Bperiodfull}{$4.1933597[4]$}
\newcommand{\KAfull}{$46.662^{+0.896}_{-0.949}$}
\newcommand{\Bqfull}{$0.848^{+0.031}_{-0.031}$}
\newcommand{\KBfull}{$55.046^{+0.944}_{-0.923}$}
\newcommand{\incdegfull}{$88.178^{+0.066}_{-0.051}$}
\newcommand{\MAfull}{$0.248^{+0.005}_{-0.004}$}
\newcommand{\MBfull}{$0.210^{+0.004}_{-0.004}$}
\newcommand{\RAfull}{$0.298^{+0.004}_{-0.005}$}
\newcommand{\RBfull}{$0.171^{+0.008}_{-0.010}$}
\newcommand{\loggAfull}{$4.884^{+0.015}_{-0.015}$}
\newcommand{\loggBfull}{$5.293^{+0.054}_{-0.042}$}
\newcommand{\semimajaufull}{$0.03922[8]$}
\newcommand{\bayesfull}{$42726.3\pm0.3$}

\newcommand{\Brrhigh}{$0.913^{+0.042}_{-0.034}$}
\newcommand{\Brsumahigh}{$0.0579[3]$}
\newcommand{\Bcosihigh}{$0.0357[4]$}
\newcommand{\Bepochhigh}{$2460054.77780[5]$}
\newcommand{\Bperiodhigh}{$4.1933598[5]$}
\newcommand{\KAhigh}{$46.143^{+0.934}_{-0.884}$}
\newcommand{\Bqhigh}{$0.831^{+0.030}_{-0.029}$}
\newcommand{\KBhigh}{$55.532^{+0.942}_{-0.893}$}
\newcommand{\incdeghigh}{$87.955^{+0.021}_{-0.022}$}
\newcommand{\MAhigh}{$0.250^{+0.005}_{-0.004}$}
\newcommand{\MBhigh}{$0.208^{+0.005}_{-0.004}$}
\newcommand{\RAhigh}{$0.255^{+0.004}_{-0.005}$}
\newcommand{\RBhigh}{$0.233^{+0.006}_{-0.005}$}
\newcommand{\loggAhigh}{$5.023^{+0.020}_{-0.016}$}
\newcommand{\loggBhigh}{$5.021^{+0.020}_{-0.022}$}
\newcommand{\semimajauhigh}{$0.03922[8]$}
\newcommand{\bayeshigh}{$42722.9\pm0.3$}

\newcommand{\Brrmid}{$0.522^{+0.019}_{-0.017}$}
\newcommand{\Brsumamid}{$0.0546[4]$}
\newcommand{\Bcosimid}{$0.0300[6]$}
\newcommand{\Bepochmid}{$2460054.77781[5]$}
\newcommand{\Bperiodmid}{$4.1933598[5]$}
\newcommand{\KAmid}{$46.583^{+0.967}_{-0.931}$}
\newcommand{\Bqmid}{$0.846^{+0.033}_{-0.031}$}
\newcommand{\KBmid}{$55.060^{+0.987}_{-0.976}$}
\newcommand{\incdegmid}{$88.281^{+0.036}_{-0.036}$}
\newcommand{\MAmid}{$0.247^{+0.005}_{-0.005}$}
\newcommand{\MBmid}{$0.209^{+0.004}_{-0.004}$}
\newcommand{\RAmid}{$0.302^{+0.002}_{-0.002}$}
\newcommand{\RBmid}{$0.158^{+0.005}_{-0.004}$}
\newcommand{\loggAmid}{$4.871^{+0.010}_{-0.010}$}
\newcommand{\loggBmid}{$5.363^{+0.025}_{-0.025}$}
\newcommand{\semimajaumid}{$0.03920[6]$}
\newcommand{\bayesmid}{$42726.0\pm0.3$}

\usepackage{amsmath}

\usepackage{multirow}

\newcommand{\qub}{Astrophysics Research Centre, School of Mathematics and Physics, Queen's University Belfast, Belfast, BT7 1NN, UK}
\newcommand{\leic}{School of Physics and Astronomy, University of Leicester, University Road, Leicester, LE1 7RH, UK}
\newcommand{\lmu}{University Observatory Munich, Faculty of Physics, Ludwig-Maximilians-Universit{\"a}t M{\"u}nchen, Scheinerstr. 1, 81679, Munich, Germany}
\newcommand{\warw}{Department of Physics, University of Warwick, Gibbet Hill Road, Coventry CV4 7AL, UK}
\newcommand{\warceh}{Centre for Exoplanets and Habitability, University of Warwick, Gibbet Hill Road, Coventry CV4 7AL, UK}
\newcommand{\mssl}{Mullard Space Science Laboratory, University College London, Holmbury St Mary, Dorking, Surrey, RH5 6NT, UK}
\newcommand{\minn}{University of Minnesota, School of Physics and Astronomy, 116 Church Street SE Minneapolis, MN 55455, USA}
\newcommand{\oxford}{Department of Physics, University of Oxford, Keble Rd, Oxford OX1 3RH, UK}
\newcommand{\adler}{Science Engagement Division, The Adler Planetarium, Chicago, IL 60605, USA}
\newcommand{\esa}{European Space Agency (ESA), European Space Research and Technology Centre (ESTEC), Keplerlaan 1, 2201 AZ Noordwijk, The Netherlands}
\newcommand{\udp}{Instituto de Estudios Astrofísicos, Facultad de Ingeniería y Ciencias, Universidad Diego Portales, Av. Ejército Libertador 441, Santiago, Chile}
\newcommand{\cata}{Centro de Excelencia en Astrofísica y Tecnologías Afines (CATA), Camino El Observatorio 1515, Las Condes, Santiago, Chile}
\newcommand{\citizen}{Planet Hunters NGTS Citizen Scientist}

\begin{document}

\title{NGTS-EB-8: A double-lined eclipsing M+M binary discovered by citizen scientists}

\correspondingauthor{Sean M. O'Brien}
\email{sean.obrien@qub.ac.uk}

\author[orcid=0000-0001-7367-1188,sname="O'Brien",gname="Sean M."]{Sean M. O'Brien}
\affiliation{\qub}
\email{sean.obrien@qub.ac.uk}

\author[orcid=0000-0003-4365-1455,sname="Schwamb",gname="Megan E."]{Megan E. Schwamb}
\affiliation{\qub}
\email{m.schwamb@qub.ac.uk}

\author[orcid=0000-0002-9718-3266,sname="Watson",gname="Christopher A."]{Christopher A. Watson}
\affiliation{\qub}
\email{c.a.watson@qub.ac.uk}

\author[orcid=0000-0002-5254-2499,sname="Nielsen",gname="Louise D."]{Louise D. Nielsen}
\affiliation{\lmu}
\email{louise.nielsen@lmu.de}

\author[orcid=0000-0001-7904-4441,sname="Bryant",gname="Edward M."]{Edward M. Bryant}
\affiliation{\warw}
\affiliation{\warceh}
\affiliation{\mssl}
\email{edward.m.bryant@warwick.ac.uk}

\author[orcid=0000-0003-2478-0120,sname="Casewell",gname="Sarah L."]{Sarah L. Casewell}
\affiliation{\leic}
\email{slc25@leicester.ac.uk}

\author[orcid=0000-0003-0684-7803,sname="Burleigh",gname="Matthew R."]{Matthew R. Burleigh}
\affiliation{\leic}
\email{mrb1@leicester.ac.uk}

\author[orcid=0000-0002-1067-8558,sname="Fortson",gname="Lucy"]{Lucy Fortson}
\affiliation{\minn}
\email{lffortson@gmail.com}

\author[orcid=0000-0002-4259-0155,sname="Gill",gname="Samuel"]{Samuel Gill}
\affiliation{\warw}
\affiliation{\warceh}
\email{samuel.gill@warwick.ac.uk}

\author[orcid=0000-0001-5578-359X,sname="Lintott",gname="Chris J."]{Chris J. Lintott}
\affiliation{\oxford}
\email{chris.lintott@physics.ox.ac.uk}

\author[orcid=0009-0004-4519-5080,sname="Hobbs",gname="Katlyn L."]{Katlyn L. Hobbs}
\affiliation{\qub}
\email{khobbs01@qub.ac.uk}

\author[orcid=0009-0004-7473-4573,sname="Apergis",gname="Ioannis "]{Ioannis  Apergis}
\affiliation{\warw}
\affiliation{\warceh}
\email{ioannis.apergis@warwick.ac.uk}

\author[orcid=0000-0001-6023-1335,sname="Bayliss",gname="Daniel"]{Daniel Bayliss}
\affiliation{\warw}
\affiliation{\warceh}
\email{d.bayliss@warwick.ac.uk}

\author[orcid=0000-0002-1416-2188,sname="Fern\'andez Fern\'andez",gname="Jorge"]{Jorge Fern\'andez Fern\'andez}
\affiliation{\warw}
\affiliation{\warceh}
\email{jorge.fernandez-fernandez.2@warwick.ac.uk}

\author[orcid=0000-0002-3164-9086,sname="G{\"u}nther",gname="Maximilian N."]{Maximilian N. G{\"u}nther}
\affiliation{\esa}
\email{maximilian.guenther@esa.int}

\author[orcid=0000-0002-8675-182X,sname="Hawthorn",gname="Faith"]{Faith Hawthorn}
\affiliation{\warw}
\affiliation{\warceh}
\email{faith.hawthorn.2@warwick.ac.uk}

\author[orcid=0000-0003-2733-8725,sname="Jenkins",gname="James S."]{James S. Jenkins}
\affiliation{\udp}
\affiliation{\cata}
\email{james.jenkins@mail.udp.cl}

\author[orcid=0009-0006-0719-9229,sname="Kendall",gname="Alicia"]{Alicia Kendall}
\affiliation{\leic}
\email{ak842@leicester.ac.uk}

\author[orcid=0000-0003-1631-4170,sname="McCormac",gname="James"]{James McCormac}
\affiliation{\warw}
\affiliation{\warceh}
\email{J.J.McCormac@warwick.ac.uk}

\author[orcid=0000-0001-6391-9266,sname="de Mooij",gname="Ernst J. W."]{Ernst J. W. de Mooij}
\affiliation{\qub}
\email{e.demooij@qub.ac.uk}

\author[orcid=0009-0009-2175-7284,sname="Rodel",gname="Toby"]{Toby Rodel}
\affiliation{\qub}
\email{trodel01@qub.ac.uk}

\author[orcid=0000-0001-8018-0264,sname="Saha",gname="Suman"]{Suman Saha}
\affiliation{\udp}
\affiliation{\cata}
\email{suman.saha@mail.udp.cl}

\author[orcid=0000-0002-1113-4122,sname="Trouille",gname="Laura"]{Laura Trouille}
\affiliation{\adler}
\email{trouille@zooniverse.org}

\author[orcid=0000-0001-6604-5533,sname="West",gname="Richard G."]{Richard G. West}
\affiliation{\warw}
\affiliation{\warceh}
\email{richard.west@warwick.ac.uk}

\author[orcid=0000-0003-1452-2240,sname="Wheatley",gname="Peter J."]{Peter J. Wheatley}
\affiliation{\warw}
\affiliation{\warceh}
\email{p.j.wheatley@warwick.ac.uk}

\author[sname="Agafitei",gname="Marius Constantin"]{Marius Constantin Agafitei}
\affiliation{\citizen}
\email{marius_aga@yahoo.com}

\author[sname="Apayd{\i}n",gname="Deniz R{\"u}zgar"]{Deniz R{\"u}zgar Apayd{\i}n}
\affiliation{\citizen}
\email{alpasalp220@gmail.com}

\author[sname="Baeten",gname="Elisabeth"]{Elisabeth Baeten}
\affiliation{\citizen}
\email{els.baeten@skynet.be}

\author[orcid=0000-0001-8731-9281,sname="Baller",gname="Bruce"]{Bruce Baller}
\affiliation{\citizen}
\email{tnballer10@gmail.com}

\author[sname="Carabott",gname="Jeff"]{Jeff Carabott}
\affiliation{\citizen}
\email{Jcarabott@sympatico.ca}

\author[sname="Chesson",gname="Sallyann"]{Sallyann Chesson}
\affiliation{\citizen}
\email{NomadPurple@live.com}

\author[sname="Freigeiro",gname="Sebasti\'{a}n Alejandro"]{Sebasti\'{a}n Alejandro Freigeiro}
\affiliation{\citizen}
\email{sebastian_freigeiro@hotmail.com}

\author[sname="Gonano",gname="Virgilio"]{Virgilio Gonano}
\affiliation{\citizen}
\email{wirg78@gmail.com}

\author[sname="Hanke",gname="Matthias"]{Matthias Hanke}
\affiliation{\citizen}
\email{matthias.hanke1@gmx.de}

\author[sname="Hermes",gname="Pete"]{Pete Hermes}
\affiliation{\citizen}
\email{petehermes@aol.com}

\author[sname="Hildebrand",gname="Avery"]{Avery Hildebrand}
\affiliation{\citizen}
\email{avery.hildebrand@gmail.com}

\author[sname="Langley",gname="John S."]{John S. Langley}
\affiliation{\citizen}
\email{jlan735436@aol.com}

\author[orcid=0009-0008-0576-556X,sname="Lim",gname="See Min"]{See Min Lim}
\affiliation{\citizen}
\email{seemin.lim@u.nus.edu}

\author[sname="McCarthy",gname="Leo Ryan"]{Leo Ryan McCarthy}
\affiliation{\citizen}
\email{importantpterosaur@gmail.com}

\author[sname="Mitchell",gname="Graham"]{Graham Mitchell}
\affiliation{\citizen}
\email{g5845mitchell@telus.net}

\author[sname="O'Neill",gname="Ken"]{Ken O'Neill}
\affiliation{\citizen}
\email{oken686@gmail.com}

\author[sname="Pearson",gname="Charles R."]{Charles R. Pearson}
\affiliation{\citizen}
\email{chuckp87.gm@gmail.com}

\author[sname="Reket",gname="Nolan"]{Nolan Reket}
\affiliation{\citizen}
\email{muisnolan@gmail.com}

\author[sname="Riethmiller",gname="Jeanne"]{Jeanne Riethmiller}
\affiliation{\citizen}
\email{deriethmiller@gmail.com}

\author[sname="Saeftel",gname="Juergen"]{Juergen Saeftel}
\affiliation{\citizen}
\email{juergen.saeftel@web.de}

\author[orcid=0000-0003-4864-5484,sname="Sainio",gname="Arttu"]{Arttu Sainio}
\affiliation{\citizen}
\email{arttu.sainio@elisanet.fi}

\author[sname="Steiner",gname="Charlie"]{Charlie Steiner}
\affiliation{\citizen}
\email{charlie.steiner68@gmail.com}

\author[sname="Strickland",gname="Amanda"]{Amanda Strickland}
\affiliation{\citizen}
\email{amanda.strickland@gmail.com}

\author[sname="Tanner",gname="Christopher"]{Christopher Tanner}
\affiliation{\citizen}
\email{c.tanner@outlook.de}

\author[orcid=0000-0002-0654-4442,sname="Terentev",gname="Ivan A."]{Ivan A. Terentev}
\affiliation{\citizen}
\email{iterentie@mail.ru}

\author[sname="Tiu",gname="Ernest Jude P."]{Ernest Jude P. Tiu}
\affiliation{\citizen}
\email{ernestjudet@gmail.com}

\author[sname="Tumanov",gname="Sergey Y."]{Sergey Y. Tumanov}
\affiliation{\citizen}
\email{sergius2013@yandex.ru}

\author[sname="Urszula",gname="Marciniak"]{Marciniak Urszula}
\affiliation{\citizen}
\email{glencora@interia.pl}

\author[sname="Vahlenkamp",gname="Pia"]{Pia Vahlenkamp}
\affiliation{\citizen}
\email{pia.vahlenkamp@t-online.de}

\author[sname="de Vroome",gname="Femke"]{Femke de Vroome}
\affiliation{\citizen}
\email{Femkedevroome@gmail.com}

\author[sname="Wantuch",gname="Pawe\l{}"]{Pawe\l{} Wantuch}
\affiliation{\citizen}
\email{wantuch@o2.pl}

\author[sname="Woodruff",gname="Timothy"]{Timothy Woodruff}
\affiliation{\citizen}
\email{twood273@comcast.net}

\begin{abstract}
We report the identification and characterization of a new binary system composed of two near-equal mass M-dwarfs. The binary NGTS-EB-8 was identified as a planet candidate in data from the Next Generation Transit Survey (NGTS) by citizen scientists participating in the Planet Hunters NGTS project. High-resolution spectroscopic observations reveal the system to be a double-lined binary. By modeling the photometric and radial velocity observations, we determine an orbital period of $4.2$\,days and the masses and radii of both stars to be $M_A=0.250^{+0.005}_{-0.004}$\,\mbox{M$_{\odot}$}, $M_B=0.208^{+0.005}_{-0.004}$\,\mbox{M$_{\odot}$}, $R_A=0.255^{+0.004}_{-0.005}$\,\mbox{R$_{\odot}$}, $R_B=0.233^{+0.006}_{-0.005}$\,\mbox{R$_{\odot}$}. We detect Balmer line emission from at least one of the stars but no significant flare activity. We note that both components lie in the fully convective regime of low-mass stars ($\lesssim0.35$\,\mbox{M$_{\odot}$}), therefore can be a valuable test for stellar evolutionary models. We demonstrate that the photometric observations, speckle imaging and initial radial velocity measurements were unable to identify the true nature of this system and highlight that high-resolution spectroscopic observations are crucial in determining whether systems such as this are in fact binaries.
\end{abstract}


\section{Introduction}\label{sec:intro}
The accurate determination of stellar parameters forms the foundation of many fields of astronomy such as: understanding star formation and evolution \citep[e.g.][]{Chabrier2007LowMass,Paxton2013MESA,Sen2022MassiveBinaries}; obtaining accurate planetary parameters \citep[e.g.][]{Hartman2015HATS6bSaturnMDwarf,Perryman2018ExoHandbook,Morales2022LateTypeEBs}; and the age-dating of stars \citep[e.g.][]{Hillenbrand2004PMSTracks,Mathieu2007PMS,David2019AgeDetermine}.
Double-lined eclipsing binaries are extremely valuable benchmark systems in this context. The analysis of their light curves and radial velocities (RVs) can provide precise, model-independent measurements of the radii and masses of the stars that can then be compared to theoretical predictions \citep[e.g.][]{Torres2010MassRadiiStars,Stassun2016EBSED,Kirk2016KeplerEBs,Prsa2022TESSEBs}.
However, in the case of M-dwarfs, there exist discrepancies between theoretical masses and radii from evolutionary models and those measured from radial velocity and photometric observations of eclipsing binaries \citep[][]{Hoxie1970VLMS,Spada2013RadiusInflation,Parsons2018MdwarfInflation}.
This `radius inflation problem,' where the observed radii of M-dwarfs are found to be larger than those predicted by models, is found for both early-type M-dwarfs and those beyond the fully convective boundary \citep[e.g.][]{Feiden2014ConvectionInhibition,Mann2015ConstrainMDwarfs,Parsons2018MdwarfInflation,Kesseli2018Inflation,Duck2023EBLMX}. A number of mechanisms have been proposed to explain this phenomenon such as: magnetic effects \citep[][]{Mullan2001Magnets,Chabrier2007LowMass,LopezMorales2007Magnetic,MacDonald2014Magnets,Morales2022LateTypeEBs}; or metallicity \citep[e.g.][]{Berger2006Metallicity,vonBoetticher2019Inflation}, however the detection and subsequent characterization of more eclipsing binaries is needed to understand this problem further.

In addition to eclipsing binaries being valuable tools to improve the accuracy of stellar parameters, it is important to identify systems that may be masquerading as exoplanets in the context of planet occurrence rates \citep[e.g.][]{Fressin2013KeplerFPandOccurrRate,Desert2015KeplerFP,Santerne2016SOPHIEKeplerGiantProperties,Triaud2017EBLM,Collins2018KELT_FPCat,TardugnoPoleo2024NotPlaNET}.
The process of confirming or ruling out planet candidates is especially important as we probe the limits of planet formation where detections are sparse and the derivation of occurrence rates often relies on including planet candidates as true detections \citep[e.g.][]{Dressing2013OccurrenceFGK,Gan2023TESSMDwarfGiantOccurrence,Bryant2023OccurrenceGiantsMDwarfs}.
The study of giant planets orbiting low-mass stars is of particular interest for planet formation models as the theory of planet formation via core-accretion predicts that giant planets should be rarer around M-dwarfs compared with FGK stars \citep{Kennedy2008CoreAccretion}. The lower protoplanetary disc mass \citep{Andrews2013PPDMassRelation} and longer dynamical timescales around low-mass stars slow the processes of planet formation \citep[e.g.][]{Laughlin2004CoreAccretionTimescale}.
There exists tension between theoretical predictions and observational results in this regime. Synthesis models from \citet{Burn2021NGPPSIV_Lowmass} suggest the occurrence rate for giant planets around stars with $M_*<0.4$\,\msun\ should be zero. Meanwhile, occurrence rates from radial velocity and transit surveys have constrained the occurrence rate of close-in giant planets orbiting low-mass stars to $0.19 - 0.5\%$ \citep{Obermeier2016APanPlanetsHJCoolDwarfOccurrence,Gan2023TESSMDwarfGiantOccurrence,Bryant2023OccurrenceGiantsMDwarfs,Mignon2025HARPSMDwarfs,Glusman2025}.
However, as shown in \citet{Bryant2023OccurrenceGiantsMDwarfs} (hereafter \citetalias{Bryant2023OccurrenceGiantsMDwarfs}), these occurrence rates can vary across an order of magnitude between the cases where all or none of the outstanding planet candidates are considered real exoplanets. Thus, the identification of which planet candidates are genuine and which are false positives is necessary for making more precise measurements of the occurrence rate in this sparse region of the parameter space.

It is in these contexts that we present the identification of \nigelid\ as an eclipsing binary system composed of two late M-dwarfs. Hereafter we refer to this system as \nigelsysname\ \citep[see][for details of this naming convention]{Rodel2025NGTSEB7}. The transit signal of this system was detected independently by \citetalias{Bryant2023OccurrenceGiantsMDwarfs} in data from the \textit{Transiting Exoplanet Survey Satellite} \citep[\tess;][]{Ricker2015TESS}, and in data from the Next Generation Transit Survey \citep[NGTS;][]{Wheatley2018NGTS} through the Planet Hunters NGTS citizen science project\footnote{\href{http://ngts.planethunters.org/}{ngts.planethunters.org}} \citep{OBrien2024PHNGTSPaper1}.
We describe the previous observations and new data obtained for \nigelsysname\ and the method applied to measure the radial velocities and characterize the spectra of the system in Section~\ref{sec:data}.
Section~\ref{sec:model} describes the modeling of all the available photometric and radial velocity data with a variety of models to derive orbital and physical parameters. We analyze the flare activity of the system in Section~\ref{sec:flares}. In Section~\ref{sec:discussion}, we discuss the properties of this system determined from the different modeling approaches in the context of other M+M double-lined eclipsing binaries and the M-dwarf radius inflation problem, and in connection with constraining the occurrence rates of giant planets orbiting low-mass stars.
Finally, in Section~\ref{sec:conclusions} we outline our conclusions and the possibility of future work on this system.

\section{Data}\label{sec:data}
Here we describe the data obtained for \nigelsysname.
We briefly describe the previous photometric and speckle imaging observations obtained for this system in Section~\ref{sec:previousdata}. A description of the additional photometry obtained from the Asteroid Terrestrial-impact Last Alert System \citep[ATLAS;][]{Tonry2018ATLAS} survey is provided in Section~\ref{sec:atlas}. The spectroscopic observations, method applied to extract radial velocity measurements and spectroscopic characterization are described in Section~\ref{sec:rvs}. 
Table~\ref{tab:catalog_info} provides information on the identifiers, coordinates and magnitudes from various sources for \nigelsysname.

\begin{table}
    \centering
    \caption{\nigelsysname\ catalog information.}\label{tab:catalog_info}
    \begin{tabular}{ccc}
    \hline
    \hline
         Property&  Value & Source\\ \hline
         \hline
         \multicolumn{3}{c}{\textit{Identifiers}} \\ 
         \hline
         2MASS ID& 11551666-4008570 & 2MASS \\
         \gaia\ Source ID&  3459141135209729536& \gaia\ DR3  \\
         TIC ID&  165227846& TIC v8.2  \\
         \hline
         \multicolumn{3}{c}{\textit{Coordinates}} \\ 
         \hline
         RA (J2000, HH:MM:SS.SSS) & 11:55:16.668 & \gaia\ DR3  \\
         Dec. (J2000, DD:MM:SS.SSS) & -40:08:57.07 & \gaia\ DR3  \\
         \hline
         \multicolumn{3}{c}{\textit{Proper motion and parallax}} \\ 
         \hline
         $\mu_{\mathrm{RA}}$ (mas y$^{-1}$)&  $-18.916\pm 0.017$& \gaia\ DR3  \\
         $\mu_{\mathrm{Dec}}$ (mas y$^{-1}$)&  $-48.513\pm0.017$& \gaia\ DR3  \\
         Parallax (mas)&  $16.04\pm0.31$& \gaia\ DR3  \\
         \hline
         \multicolumn{3}{c}{\textit{Magnitudes}} \\ 
         \hline
         V (mag)&  $16.365\pm 1.133$& UCAC4  \\
         B (mag)& $18.372\pm 0.161$& 2MASS  \\
         g (mag)& $16.640\pm0.009$& SkyMapper  \\
         r (mag)& $15.627\pm0.006$& SkyMapper  \\
         i (mag)& $13.890\pm0.004$& SkyMapper  \\
         z (mag)& $13.216\pm0.007$& SkyMapper  \\
         G (mag)& $14.7767\pm0.0005$&\gaia\ DR3\dag  \\
         BP (mag) & $16.477\pm0.004$&\gaia\ DR3\dag  \\
         RP (mag) &$13.5299\pm0.0007$&\gaia\ DR3\dag  \\
         \tess\ (mag)& $13.458\pm0.007$&TIC v8.2  \\
         J (mag)& $11.743\pm 0.021 $&2MASS  \\
         H (mag)& $11.169\pm 0.022$&2MASS  \\
         K (mag)& $10.885\pm0.019$&2MASS  \\
         W1 (mag)& $10.723\pm 0.023$&ALLWISE  \\
         W2 (mag)& $10.545\pm 0.021$&ALLWISE  \\
         \hline
    \end{tabular}
    \begin{flushleft}
        \footnotesize{Sources: 2MASS \citep{Skrutskie20062MASS}; Gaia DR3 \citep{Gaia2016GaiaMission,Gaia2023GaiaDR3}; TIC v8.2 \citep{Stassun2019RevisedTIC}; UCAC4 \citep{Zacharias2013UCAC4}; SkyMapper \citep{Onken2024SkyMapperDR4}; ALLWISE \citep{Wright2010WISE}. \dag \gaia\ magnitudes are calculated from the measured flux values, thus \gaia\ DR3 does not report magnitude uncertainties as the transformation from fluxes to magnitudes is non-linear and would results in an asymmetric error distribution \citep{Riello2021GaiaEDR3Photometry}. We calculate the upper and lower magnitude uncertainties as $\Delta m_{\mathrm{upper}} = -2.5 \log_{10}(\frac{f}{f + \Delta f})$ and $\Delta m_{\mathrm{lower}} = -2.5 \log_{10}(\frac{f - \Delta f}{f})$, respectively. In practice, the uncertainties are symmetric to the precision quoted.}
    \end{flushleft}
\end{table}

\subsection{Previous Observations}\label{sec:previousdata}
\citet{OBrien2024PHNGTSPaper1} presented photometry from: NGTS; four individual transit observations in different filters ($z^{\prime}$, $g^{\prime}$, $i^{\prime}$, $V$) using the Sutherland High-speed Optical Camera \citep[SHOC;][]{Coppejans2013SHOC} at the South African Astronomical Observatory (SAAO); and four \tess\ sectors (Sectors 10, 37, 63 and 64). The \tess\ Sector 10 data was also presented in the discovery of this system by \citetalias{Bryant2023OccurrenceGiantsMDwarfs}.
Gemini/Zorro speckle imaging \citep{Scott2021Zorro,Howell2022Zorro} was also presented in \citet{OBrien2024PHNGTSPaper1} that revealed no stellar companions within 1.17\,arcsec of the target at approximately the 4–7\,mag limit in either Zorro filter. We refer the reader to Section~2 and Section~5.1 of \citet{OBrien2024PHNGTSPaper1} for details of these observations. From the photometric observations, \citet{OBrien2024PHNGTSPaper1} measured an orbital period of $P=2.09$\,days and a transit depth of 13.1\% that, when combined with the stellar radius of 0.32\,\rsun\ from the \tess\ Input Catalog \citep[TIC;][]{Stassun2018TICI,Stassun2019RevisedTIC,Paegert2021TICv8.2}, results in an estimated companion radius of $1.61^{+0.13}_{-0.13}$\,\rjup.
We also note that the \tess\ Data Validation Report \citep{Twicken2019TESSDataValidation} generated due to the automated detection of transit events in \tess\ data did not flag this candidate as a planet candidate or an eclipsing binary. We analyze these data and the additional follow-up data described below to characterize this system.

\subsection{ATLAS Photometry}\label{sec:atlas}
The primary mission of ATLAS \citep{Tonry2018ATLAS} is to detect potentially hazardous near-Earth asteroids by conducting an all-sky survey. The survey consists of four 0.5m telescopes with two units in Hawai`i, Haleakala (HKO) and Mauna Loa (MLO), and one each in El Sauce, Chile (CHL) and Sutherland, South Africa (STH). This provides coverage of 80\% of the night sky every night. The telescopes reach a limiting magnitude of approximately 19.5\,mag in two nonstandard wide-band filters, cyan ($c$, 420-650\,nm) and orange ($o$, 560-820\,nm). Given our target has a declination of approximately $-40$\,deg and median magnitudes of $m_o\mysim$14.7\,mag and $m_c\mysim$16.3\,mag, we are able to access the existing ATLAS data to obtain a long baseline (9\,years) of photometry for \nigelsysname. These data allow the opportunity to characterize the long-term flare activity of the star (Section~\ref{sec:flares}). We note that frequent observations began in late 2017 following the installation of the second telescope (MLO) in Hawai`i, while the two southern hemisphere telescope units (CHL and STH) came online in late 2021, providing even higher cadence observations for this southern target.
We obtain reduced ATLAS photometric data using the ATLAS Forced Photometry Server\footnote{\url{https://fallingstar-data.com/forcedphot/}} \citep{Shingles2021ATLASForcedPhot}. Further details of the ATLAS system and data-reduction pipeline are described in \citet{Tonry2018ATLAS, Tonry2018ATLASRefCat} and \citet{Smith2020ATLASTSS}.
We clean the ATLAS data to remove bad measurements, following a method similar to that described in \citet{Dobson2023ATLASPhaseCurves}, by removing observations where the apparent magnitude is dimmer than (i) the $5\sigma$ limiting magnitude of the image (to ensure the observation was of good quality), and (ii) the $3\sigma$ upper magnitude limit derived from the flux uncertainty (ensuring the target object could be detected on the image). We also remove data points with uncertainties larger than the 95th percentile of all uncertainties to remove observations that were likely taken during poor observing conditions.

To further mitigate against telescope systematics, we generate ATLAS light curves for a sample of nearby stars and then perform relative photometry.
We select all stars from the TIC \citep{Stassun2018TICI,Stassun2019RevisedTIC} within 3\,arcmin of our target and with $13 \leq $\,\grp\,$\leq 16$ where \grp\ is the \gaia\ RP magnitude \citep{Gaia2016GaiaMission,Gaia2023GaiaDR3}. We choose this magnitude range to ensure that the comparison stars are of similar brightness to our target (\grp = 13.53\,mag), while also not saturating the ATLAS detector or being too faint to generate a good quality light curve. We select the 3\,arcmin limit to provide enough comparison stars to construct a reliable master reference `star' while ensuring the comparison stars are close enough to the target that they may exhibit any systematics effecting all stars in the field. This provides a sample of 15 comparison stars. We clean these light curves using the same procedure as described above.
We divide the target star flux by the mean of the fluxes of the comparison stars and normalize the light curve by dividing by the median flux.
We remove measurements for the target star where there are less than 15 comparison stars at the given timestamp.
Prior to this relative photometry step, the measurements from the CHL unit show significant scatter ($\sigma=0.074$) compared with those measured by the STH unit during the same time period ($\sigma=0.025$). We note that the relative photometry method significantly reduces the scatter in the measurements made by the CHL unit ($\sigma=0.023$).

\begin{figure}
    \centering
    \includegraphics[width=0.95\columnwidth]{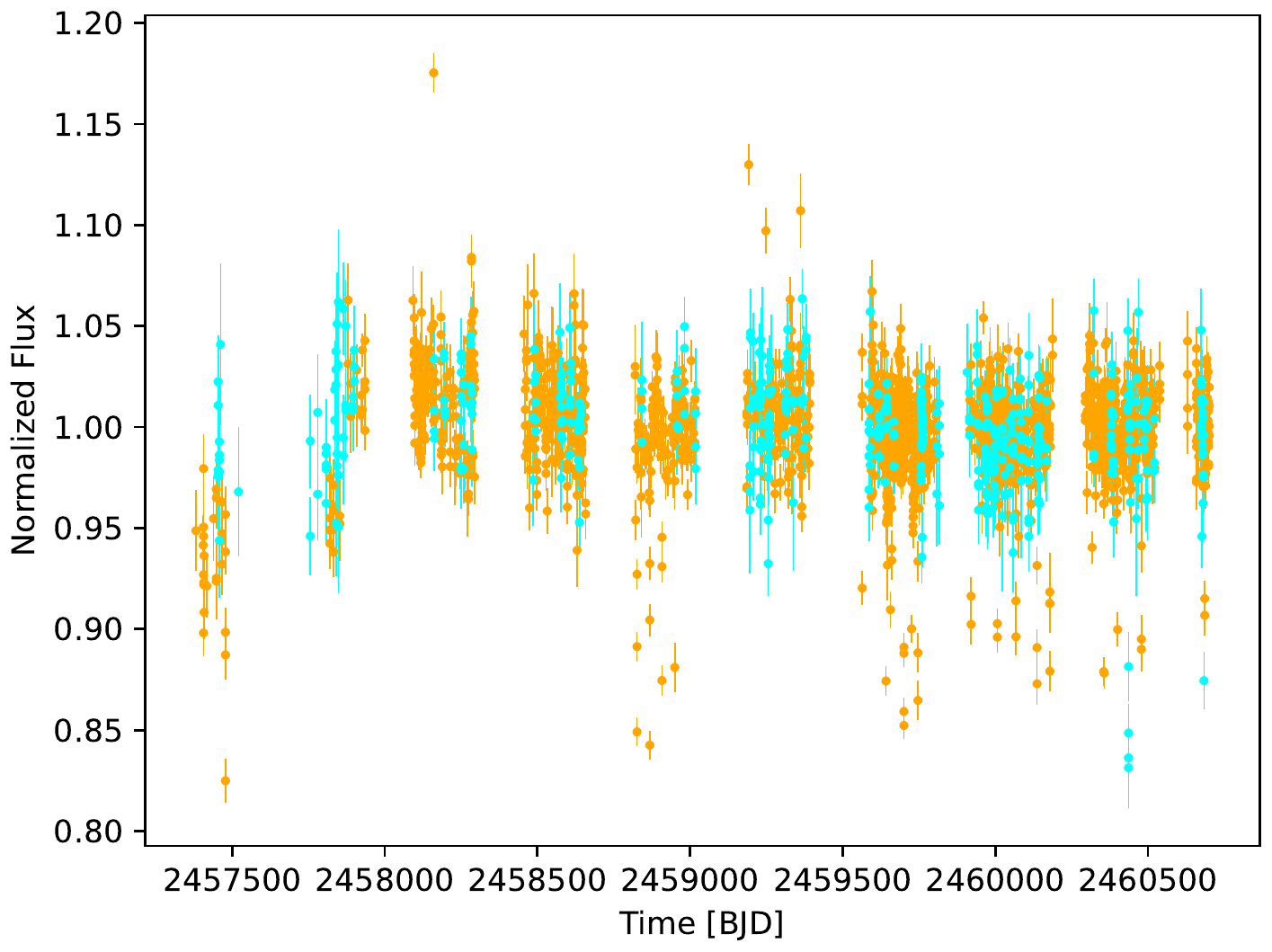}
    \caption{ATLAS photometric data for \nigelsysname. $o$-band and $c$-band measurements are shown in orange and cyan, respectively.}
    \label{fig:atlas_lc}
\end{figure}

Figure~\ref{fig:atlas_lc} shows the ATLAS $o$- and $c$-band light curves after performing relative photometry.
We can tentatively attribute the flux measurements that are significantly below the median to be due to the primary and secondary eclipses. Figure~\ref{fig:alles_photom} shows the phase folded light curves for both ATLAS bandpasses where we can determine that we do indeed recover the primary and secondary eclipses in the ATLAS light curves.
The measurement of the eclipses in ATLAS data presents the possibility of utilizing all-sky surveys such as ATLAS, the Zwicky Transient Facility \citep[ZTF;][]{Bellm2019ZTF}, the Panoramic Survey Telescope and Rapid Response System \citep[Pan-STARRS][]{Chambers2016PanSTARRS}, and, in the near future, the Legacy Survey of Space and Time \citep[LSST;][]{Ivezic2019LSST}, to carry out more focused searches to detect deep transits resulting from eclipsing binaries or giant planets orbiting low-mass stars. White dwarf+main sequence binaries that also display these deep eclipses have been successfully recovered in Catalina and ZTF survey data \citep[e.g.][]{Parsons2013CatalinaWDs,Parsons2015SDSSCatalinaWDs,Brown2023ZTFWDs}. The regular, deep, long-baseline photometry of these surveys compared with conventional exoplanet transit surveys raises the possibility to probe longer period EBs and giant planets orbiting low-mass stars.

\subsection{Spectroscopy}\label{sec:rvs}
Here we describe the spectroscopic follow-up data obtained for the \nigelsysname\ system. We obtained spectroscopic observations from the Near Infra-Red Planet Searcher \citep[NIRPS;][]{Bouchy2017NIRPS,Wildi2022NIRPSFirstLight} in April 2023 and Gemini High-resolution Optical SpecTrograph \citep[GHOST;][]{Ireland2014GHOST,McConnachie2022GHOSTNatAs,Hayes2023GHOSTCommissioning,McConnachie2024GHOSTSciencePerformance} in May 2024. The radial velocities measured from the GHOST data informed the radial velocity extraction method applied to the NIRPS data. The radial velocities measured from both data informed our spectroscopic characterization of the two stars.

\subsubsection{GHOST}\label{sec:ghost}
The GHOST instrument, installed on the Gemini South telescope, provides high-resolution spectra across a wide wavelength range (383-1000\,nm). The spectra are split into a blue and a red camera with wavelength ranges of 347-542\,nm and 520-1060\,nm, respectively.
We obtained two consecutive exposures with exposure times of 700\,seconds on 2024 May 3 using the high-resolution ($R\,\mysim\,76000$) mode for GHOST. We elected to obtain two exposures to ensure that in the event of cosmic rays or other losses, we would have sufficient data and signal-to-noise to carry out the desired analysis. We used the default medium and slow readout speeds for the red and blue cameras, respectively. We determined there was little risk of object merging as our target is in an uncrowded field, therefore we elected for 1\,x\,4 (spectral\,x\,spatial) binning to improve the signal-to-noise of our observations. We obtained a telluric standard of the A0V star HIP\,60985 with an exposure time of 120\,s immediately following the science target exposures in order to mitigate the effects of the Earth's atmosphere in the data reduction/analysis.
The data were reduced using the \dragons\ software \citep{Labrie2023DRAGONSOverview,Labrie2023DRAGONSZenodo}, following the default parameters and standard steps recommended for the GHOST Data Reduction \citep{Placco2024GHOSTReduced}. These steps are bias subtraction, flat correction, wavelength calibration, sky subtraction, and correcting for barycentric motion.

\subsubsection{NIRPS}\label{sec:nirps}
The Near Infra-Red Planet Searcher \citep[NIRPS;][]{Bouchy2017NIRPS,Wildi2022NIRPSFirstLight} is a high-resolution \'{e}chelle spectrograph mounted on the ESO 3.6\,m telescope at La Silla Observatory. When combined with the High Accuracy Radial velocity Planet Searcher \citep[HARPS;][]{Mayor2003HARPS}, this allows simultaneous spectral coverage in the NIR (971 to 1854\,nm) and VIS (378 to 691\,nm) channels. We obtained HARPS spectra in addition to the NIRPS spectra described below, however the HARPS data were determined to have too low signal-to-noise and were discarded from further analysis. For \nigelsysname, we obtained 18 spectra (across 9 individual nights) with NIRPS between 2023 April 1 and 2023 April 27 under program number 111.254E.001. We obtained two consecutive exposures per night using NIRPS with exposure times ranging from 600 to 900\,s. We elected to use two sub-exposures as exposure times greater than 1000\,s are not recommended for NIRPS. This is due to the possibility of smearing of telluric absorption lines and increased readout noise \citep{NIRPS2022}. The exposure times were increased after the first three nights of observation to improve signal-to-noise following preliminary checks of the data.
NIRPS observations were obtained in High Efficiency (HE; $R\,\mysim\,75000$) mode. Data were reduced using the standard NIRPS DRS pipeline.

\subsubsection{Radial Velocity Measurement}\label{sec:rv_meas}
\begin{figure}
    \centering
    \includegraphics[width=0.9\columnwidth]{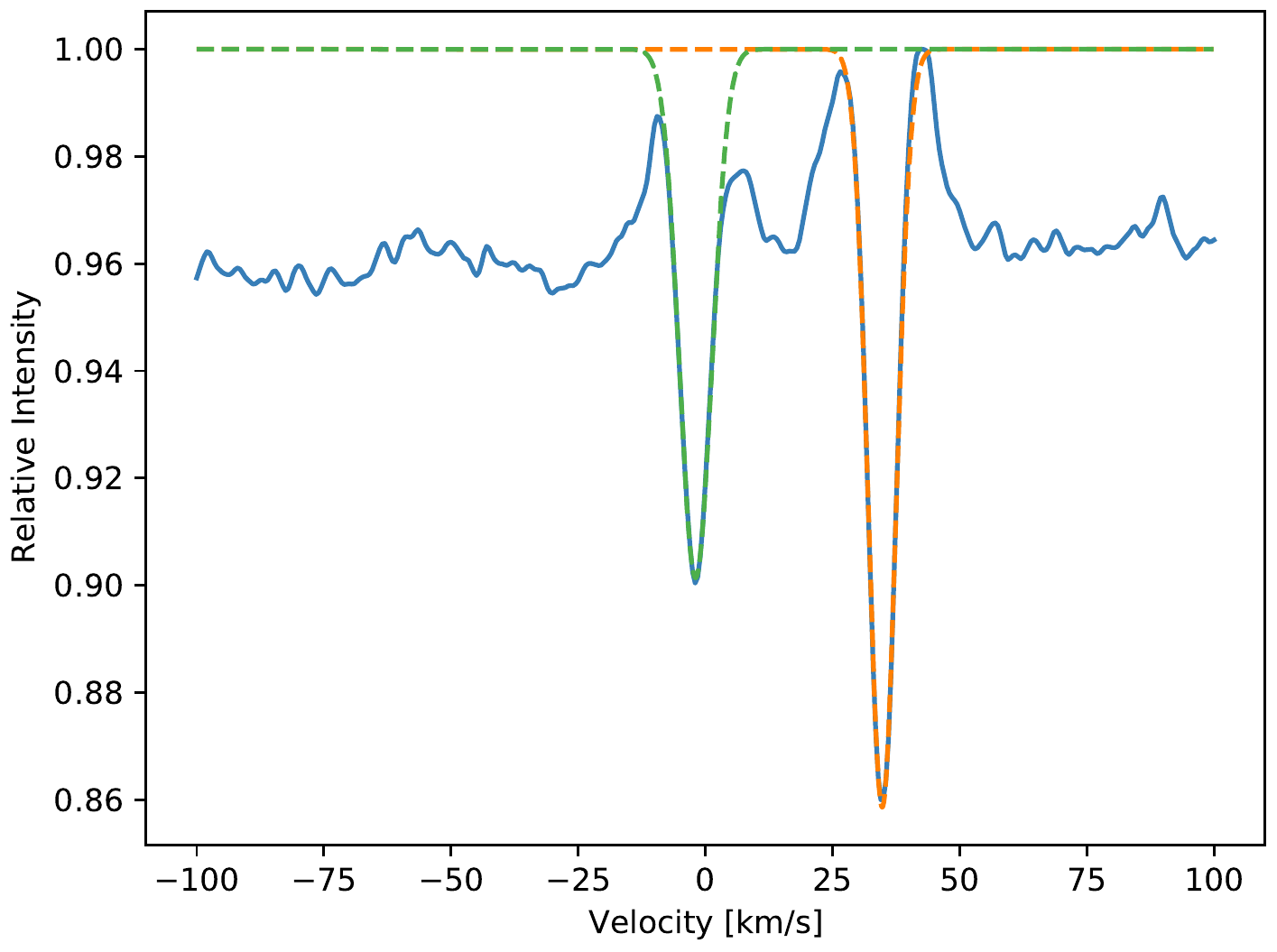}
    \caption{CCF for the first GHOST spectrum of \nigelsysname\ cross-correlated with the HARPS M5 mask. The solid blue line shows the median CCF. The orange and green dashed lines show the Gaussian fits for the primary and secondary component, respectively}
    \label{fig:ghost_ccf}
\end{figure}

We first analyze the GHOST spectra to measure the radial velocities. We mask out the telluric regions manually after visual inspection of the telluric standard spectrum. We use the \ispec\ package \citep{BlancoCuaresma2014iSpec,BlancoCuaresma2019iSpec} to fit a spline to the continuum of each spectrum and then divide through by the fitted continuum to normalize the spectra.
We then use \ispec\ to cross-correlate both spectra with an M5 stellar mask from the High Accuracy Radial velocity Planet Searcher \citep[HARPS;][]{Pepe2002HARPS,Mayor2003HARPS} pipeline to obtain a cross-correlation function \citep[CCF;][]{Baranne1996ELODIE,Pepe2002HARPS}. We opt for an M5 mask as this is the closest match of the templates available in \ispec\ to the spectral type of the presumed host star ($\mysim$M3). We calculate the CCF across a velocity range of -100 to 100\,\kms\ with step sizes of 0.5\,\kms. While the wavelength range of the mask (400-687\,nm) is much smaller than the spectral range of GHOST, this CCF is sufficient to identify the double-lined nature of the spectra (SB2). The CCF for the first exposure in Figure~\ref{fig:ghost_ccf} shows two clear troughs and we measure radial velocities of $34.77\pm0.03$\,\kms\ and $-1.93\pm0.05$\,\kms\ for the two components by fitting Gaussians to each trough. The Gaussians are fitted using \texttt{scipy.optimize.curve\_fit} using the Trust Region Reflective minimization algorithm that is designed to provide a robust fit \citep{Virtanen2020Scipy}. We measure the radial velocity as the center of the Gaussian with the errors calculated as $\sqrt{\sigma^2}$ where $\sigma^2$ is the variance on the parameter estimated by \texttt{curve\_fit}. We find velocities of $35.73\pm0.03$\,\kms\ and $-2.83\pm0.05$\,\kms\ from the second science spectrum. This confirms the double-lined nature of the spectra, revealing this system to be a binary star system.

We use the two pairs of radial velocities measured from the GHOST CCFs to obtain initial estimates for the masses of the two components that we use for subsequent analysis. This is achieved by generating model radial velocity curves for both components across a range of masses ($0.05\leq M_A,M_B \leq 0.4\,$\msun). We test two orbital periods: $P=2.10$\,days that was measured from transit photometry and reported in \citet{OBrien2024PHNGTSPaper1}; and $P=4.19$\,days, that is, twice the reported orbital period assuming the primary and secondary eclipses were erroneously attributed to being the same transit event.
We calculate the expected RV of both components at the two GHOST epochs from simple Keplerian models and compare these expected RVs to the RVs measured from the GHOST CCFs via a chi-squared statistic.
Taking the minimum chi-square value for these fits, we find better agreement with an orbital period of $P=4.19$\,days and estimate the two masses to be $\mysim0.21$\,\msun\ and $\mysim0.18$\,\msun.

\begin{figure}
    \centering
    \includegraphics[width=0.9\columnwidth]{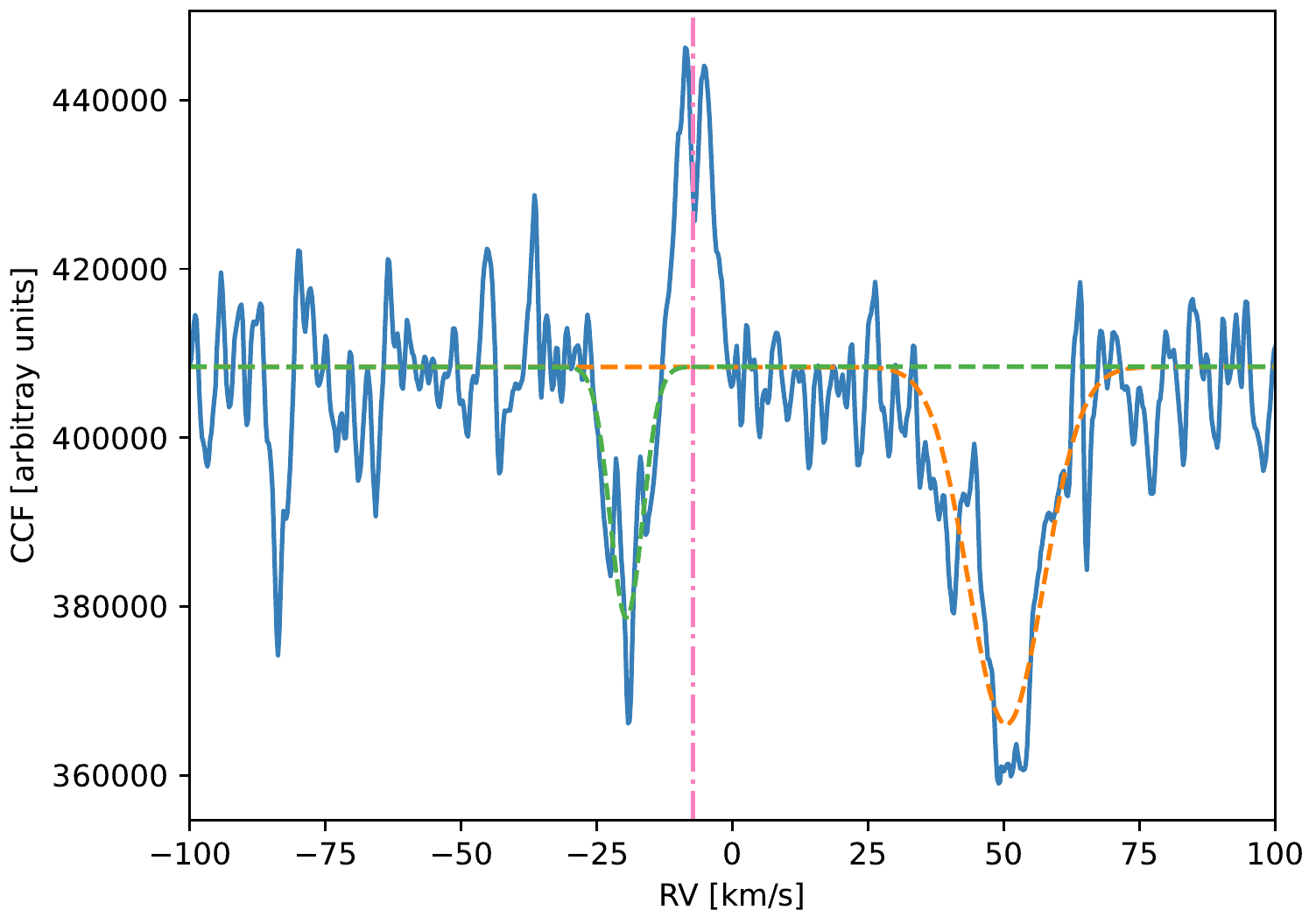}
    \caption{The CCF for the NIRPS spectrum of \nigelsysname\ from the night of 2023 April 27 cross-correlated with the NIRPS M2 mask. The solid blue line shows the CCF. The vertical dash-dot pink line shows the barycentric Earth radial velocity (BERV) value that coincides with the sky emission peak. The orange and green dashed lines show the Gaussian fits for the primary and secondary component, respectively.}
    \label{fig:nirps_ccf}
\end{figure}

Following the observations taken with GHOST and the identification of the system as an SB2, we analyzed these NIRPS data with a wide velocity window of (\vsys-100, \vsys+100)\,\kms\ when calculating the CCF. The NIRPS CCFs were calculated using an M2 mask as this is the closest match of the templates available in the NIRPS pipeline to the spectral type of the presumed stars. This reveals radial velocity signals for both components in the NIRPS CCFs and we highlight this as an example where wide velocity ranges are necessary for cross-correlation to ensure the signals of multiple stellar components are not missed \citep[e.g.][]{Standing2023BEBOP1,Sebastian2024BEBOP1}.
We utilize the mass estimates obtained from the analysis of the GHOST RVs and an orbital period of $P=4.19$\,days to calculate the expected radial velocities of each component at each NIRPS epoch. We then use these estimates as initial guesses to fit two Gaussians to the NIRPS CCFs following the same approach used for measuring the RVs from the GHOST CCFs. Figure~\ref{fig:nirps_ccf} shows the CCF for a single epoch of NIRPS data.
We extract radial velocities for both components at most epochs, although we are unable to measure the radial velocity when the signal lies close to the sky emission peak around 0\,\kms\ that persists even after data reduction.
The radial velocities and associated errors for the NIRPS and GHOST observations are listed in Table~\ref{tab:nirps_rvs}.
\begin{table*}
    \centering
    \caption{Spectroscopic data for \nigelsysname.}\label{tab:nirps_rvs}
    \begin{tabular}{rcccccc}
    \hline
        Time (BJD) & $RV_A$ (\kms) & $\sigma_{RV_A}$ (\kms)& $RV_B$ (\kms) & $\sigma_{RV_B}$ (\kms) & Inst. & $t_{\text{exp}} (s)$\\ \hline
        2460036.5838657410 & 54.02 & 1.15 & -34.74 & 1.64 & NIRPS & 600 \\
        2460036.5909606484 & 56.69 & 1.03 & -33.18 & 1.71 & NIRPS & 600 \\
        2460037.5748148146 & 45.86 & 1.01 & -37.57 & 2.63 & NIRPS & 600 \\
        2460037.5819097220 & 45.41 & 1.10 & -37.18 & 3.02 & NIRPS & 600 \\
        2460038.7501967590 & -28.48 & 1.44 & 62.72 & 1.17 & NIRPS & 600 \\
        2460038.7572916667 & -25.49 & 0.72 & 63.86 & 1.00 & NIRPS & 600 \\
        2460043.6322337960 & -20.47 & 1.64 & 61.59 & 1.81 & NIRPS & 900 \\
        2460043.6427430560 & -19.75 & 1.41 & 60.38 & 0.95 & NIRPS & 900 \\
        2460044.6956828700 & 44.27 & 0.18 & -13.36 & 0.29 & NIRPS & 900 \\
        2460044.7061921298 & 45.34 & 0.16 & -14.16 & 0.36 & NIRPS & 900 \\
        2460056.7382870370 & - & - & 29.69 & 2.88 & NIRPS & 900 \\
        2460056.7488078703 & - & - & 30.99 & 2.71 & NIRPS & 900 \\
        2460057.6154513890 & 58.36 & 0.98 & -40.42 & 2.25 & NIRPS & 900 \\
        2460057.6259722220 & 60.25 & 0.79 & -37.38 & 1.46 & NIRPS & 900 \\
        2460058.5633217595 & 42.76 & 1.06 & - & - & NIRPS & 900 \\
        2460058.5738310185 & 42.50 & 1.04 & - & - & NIRPS & 900 \\
        2460061.5710185184 & 48.29 & 0.35 & -19.59 & 0.35 & NIRPS & 900 \\
        2460061.5815277780 & 50.40 & 0.30 & -19.58 & 0.28 & NIRPS & 900 \\
        2460434.5235763890 & 34.77 & 0.03 & -1.93 & 0.05 & GHOST & 700 \\
        2460434.5362847224 & 35.73 & 0.03 & -2.83 & 0.05 & GHOST & 700 \\
        \hline
    \end{tabular}
\end{table*}

In order to determine a more accurate mass ratio to inform our spectroscopic characterization (Section~\ref{sec:spec_char}) and model fitting (Section~\ref{sec:model}), we perform a preliminary fit of the NIRPS and GHOST RVs for the primary and secondary component using the \allesfitter\ package \citep{Guenther2019allesfittercode,Guenther2021allesfitterpaper}.
\texttt{Allesfitter} combines \texttt{ellc} \citep[light curve and RV models;][]{Maxted2016ellc}, \texttt{emcee} \citep{ForemanMackey2013emcee}, \texttt{dynesty} \citep[Nested Sampling (NS);][]{Speagle2020dynesty} and \texttt{celerite} \citep{ForemanMackey2017celerite} to fit a range of data, with models available for a variety of signals including eclipsing binaries. We estimate the mass ratio of the system as $q=0.86\pm0.04$ that we then use to constrain the parameter space of models considered in Section~\ref{sec:spec_char}.

\subsubsection{Spectroscopic Characterization}\label{sec:spec_char}
We analyze the first GHOST spectrum to estimate the stellar atmospheric parameters (effective temperature (\teff), surface gravity (\logg) and metallicity (\feh)) of both components.
We did not observe a spectrophotometric standard star to perform absolute flux calibration of the spectrum of \nigelsysname\ and therefore we opt to compare our spectrum to synthetic spectra across smaller regions of the spectra. We use a selection of regions defined in the \texttt{PyHammer} software package \citep{Kesseli2017SpectralTemplates,Roulston2020PyHammer2} that is based on the \texttt{Hammer} code described in \citet{Covey2007TheHammer}. These packages are primarily designed for spectral typing stars using medium resolution ($R\mysim1000 - 10000$) spectra and thus utilize the measurement of spectral indices to determine the closest match template spectra. However, given the high-resolution ($R\mysim76000$) of the GHOST spectrum we obtained, we opt to use \chisq-fitting to more accurately compare the shape of specific absorption features in our observed spectrum to those in templates that we create from PHOENIX synthetic spectra \citep{Husser2013PhoenixV2}\footnote{\href{https://phoenix.astro.physik.uni-goettingen.de/}{https://phoenix.astro.physik.uni-goettingen.de/}}.
The regions of the spectra that we analyze are defined in Table~\ref{tab:spec_regions}.
\begin{table}
\centering
\caption{Spectral features derived from spectral indices described in \citet{Roulston2020PyHammer2}. The wavelength range defines the full range over which the \chisq-statistic is calculated. The flux normalization range defines the region used to normalize both the observed and template spectra by dividing by the median flux value in this range.}\label{tab:spec_regions}
\begin{tabular}{ccc}
\hline
Spectral Feature   & Wavelength Range (nm) & Flux Normalization Range (nm) \\ \hline
NaD              & 587.1630-594.6645     & 591.1638-593.6645            \\
Ca I          & 603.0000-618.6709     & 604.5400-606.0700            \\
TiO             & 703.3942-714.6967     & 704.3942-704.7943             \\
VO           & 735.2025-756.2082     & 742.2044-747.2057             \\
VO           & 790.3500-812.2200     & 796.2190-800.2200             \\
Na I              & 814.3241-821.3255     & 815.3241-817.5248             \\
TiO             & 837.0000-844.8000     & 840.1000-841.0000             \\
Ca II         & 847.5331-855.5347     & 851.6000-853.8000             \\
Ca II/Fe I & 861.7369-870.0000     & 862.7369-865.2376             \\
FeH              & 981.0000-998.0000     & 982.0000-986.0000             \\ \hline
\end{tabular}
\end{table}
We define the full wavelength range around each feature over which we calculate the \chisq-statistic and a flux normalization range where we take the median of each spectra in this region and divide the flux measurements across the full region to locally normalize the observed spectra and the templates.
The templates are constructed using PHOENIX synthetic spectra with $2300\leq$\teff$\leq4000$\,K, $4.00\leq$\logg$\leq5.50$ and $-4.0\leq$\feh$\leq+1.0$. We create template SB2 spectra by combining the synthetic spectra in pairs. The fluxes of each synthetic spectra are added together with equal weighting after shifting each component by the radial velocities measured in Section~\ref{sec:rv_meas} for the first GHOST spectrum. We interpolate the template SB2 spectra onto the wavelength grid of the first GHOST spectrum.

\begin{figure}
    \centering
    \includegraphics[width=0.95\linewidth]{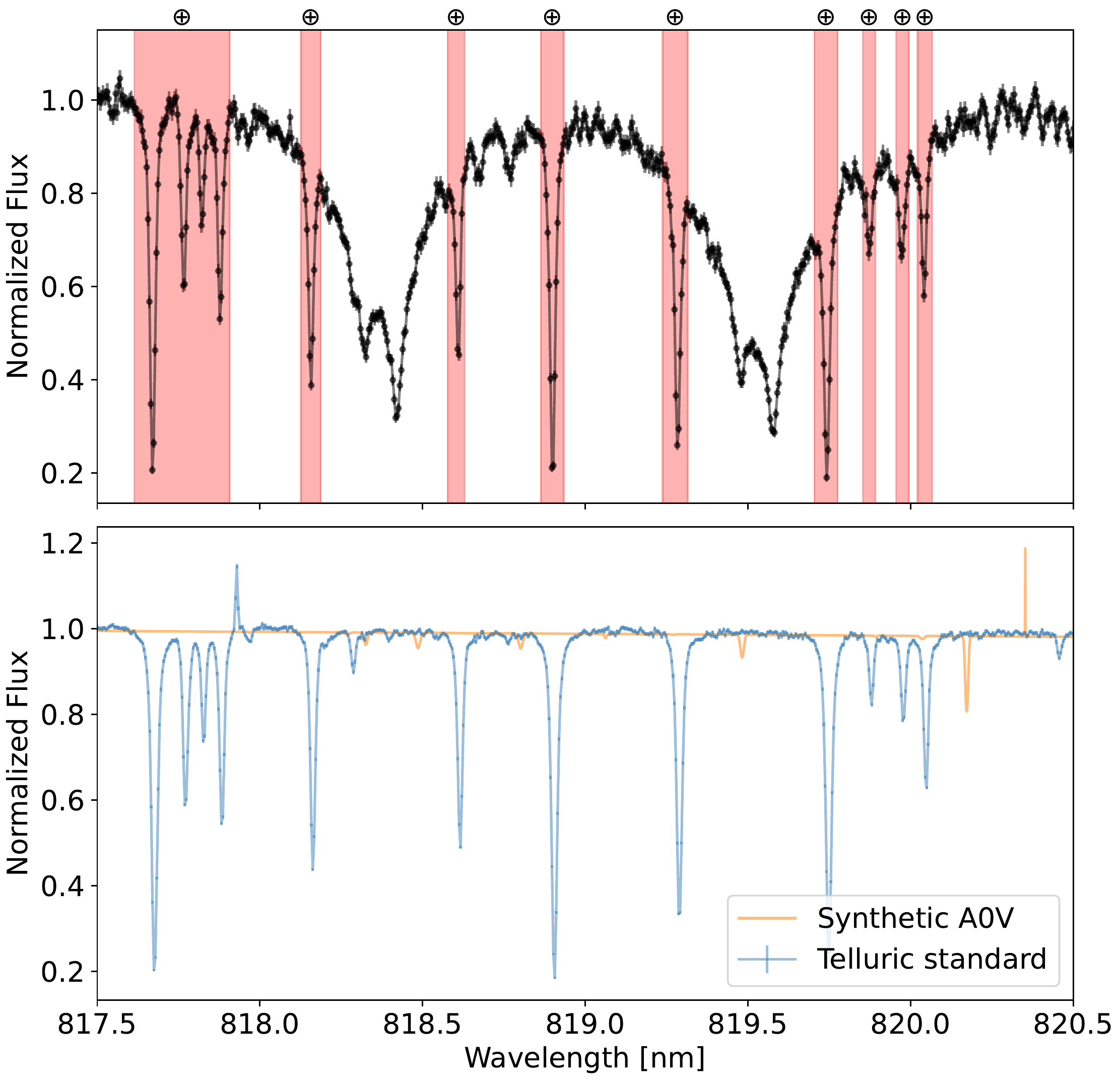}
    \caption{Telluric line identification for the Na I region, zoomed on the Na I feature around 819\,nm. (Top panel) GHOST spectrum of \nigelsysname\ shown in black. Regions identified as telluric absorption features are highlighted in red and marked with $\oplus$. (Bottom panel) GHOST spectrum of the telluric standard star (HIP\,60985) in blue and the synthetic PHOENIX spectrum that represents an A0V star in orange.}
    \label{fig:telluric_masking}
\end{figure}

\begin{figure}
    \centering
    \includegraphics[width=0.95\linewidth]{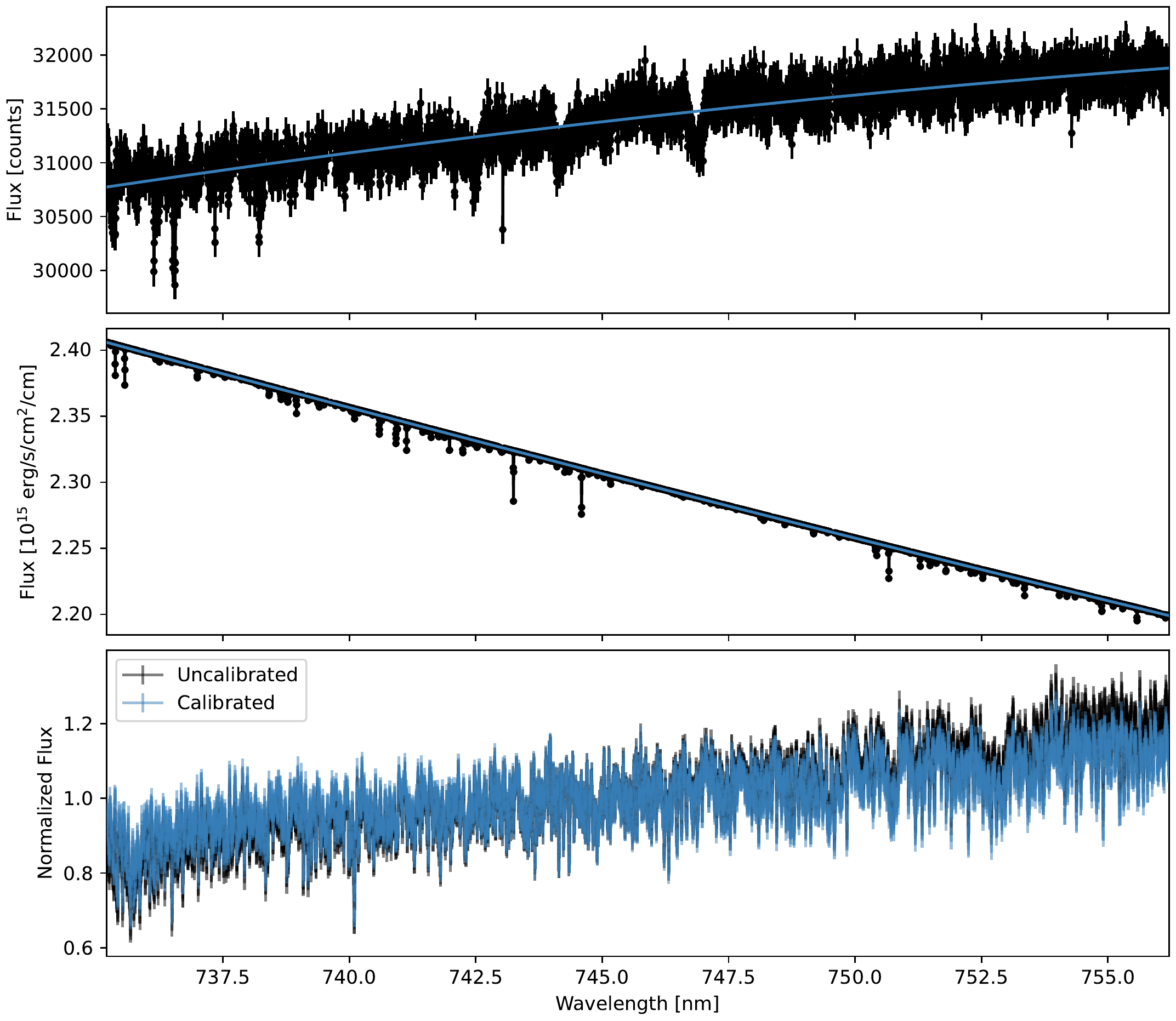}
    \caption{Relative flux calibration process for the VO feature around 746\,nm. (Top panel) GHOST spectrum of telluric standard star in black with the continuum fit shown in blue. (Middle panel) PHOENIX synthetic spectrum with \teff$=10000$\,K, \logg$=3.50$ and \feh$=-0.5$ shown in black with the continuum fit shown in blue. (Bottom panel) GHOST spectrum of \nigelsysname\ shown in black prior to the relative flux calibration and shown in blue following the relative flux calibration.}
    \label{fig:flux_cal}
\end{figure}

Prior to calculating the \chisq\ in each region, we manually remove wavelengths/pixels in the observed spectrum that show significant telluric absorption in each region defined in Table~\ref{tab:spec_regions}. We do this by visually comparing the observed spectrum of \nigelsysname\ to the spectrum of the telluric standard star and a PHOENIX model with \teff$=10000$\,K, \logg$=3.50$ and \feh$=-0.5$ that represents a synthetic (telluric-free) spectrum of an A0V star. Figure~\ref{fig:telluric_masking} shows an example of the telluric absorption lines that were identified and removed around the Na I feature.
Additionally, we perform a relative flux calibration in each region using the telluric standard star. We fit the continua of the telluric standard star and the PHOENIX model with \teff$=10000$\,K, \logg$=3.50$ and \feh$=-0.5$ using a 2nd-order polynomial fit to the continuum regions. We calculate the ratio of these continua to approximate the effects of the instrument as a function of wavelength and then divide our spectrum of \nigelsysname\ by these values. Figure~\ref{fig:flux_cal} shows an example of this relative flux calibration step for the VO feature around 746\,nm. We see that this step partially removes the gradient in this region of the spectrum that is introduced by the instrument.
We compare the calibrated spectrum of \nigelsysname\ to each template spectra by calculating the \chisq\ across all regions defined in Table~\ref{tab:spec_regions}. The number of data points across all regions is $N_{\text{data}} = 30414$. We do not fit directly for \teff, \logg\ and \feh\ but rather compare our observed spectrum to a fixed grid of template spectra and therefore the number of degrees of freedom is $\nu = N_{\text{data}} = 30414$. Given the number of degrees of freedom is consistent between the fits to all the template spectra, we quote reduced \chisq\ values ($\chi^2_{\nu} = \frac{\chi^2}{\nu}$) in the following discussion.

\begin{figure}
    \centering
    \includegraphics[width=0.95\columnwidth]{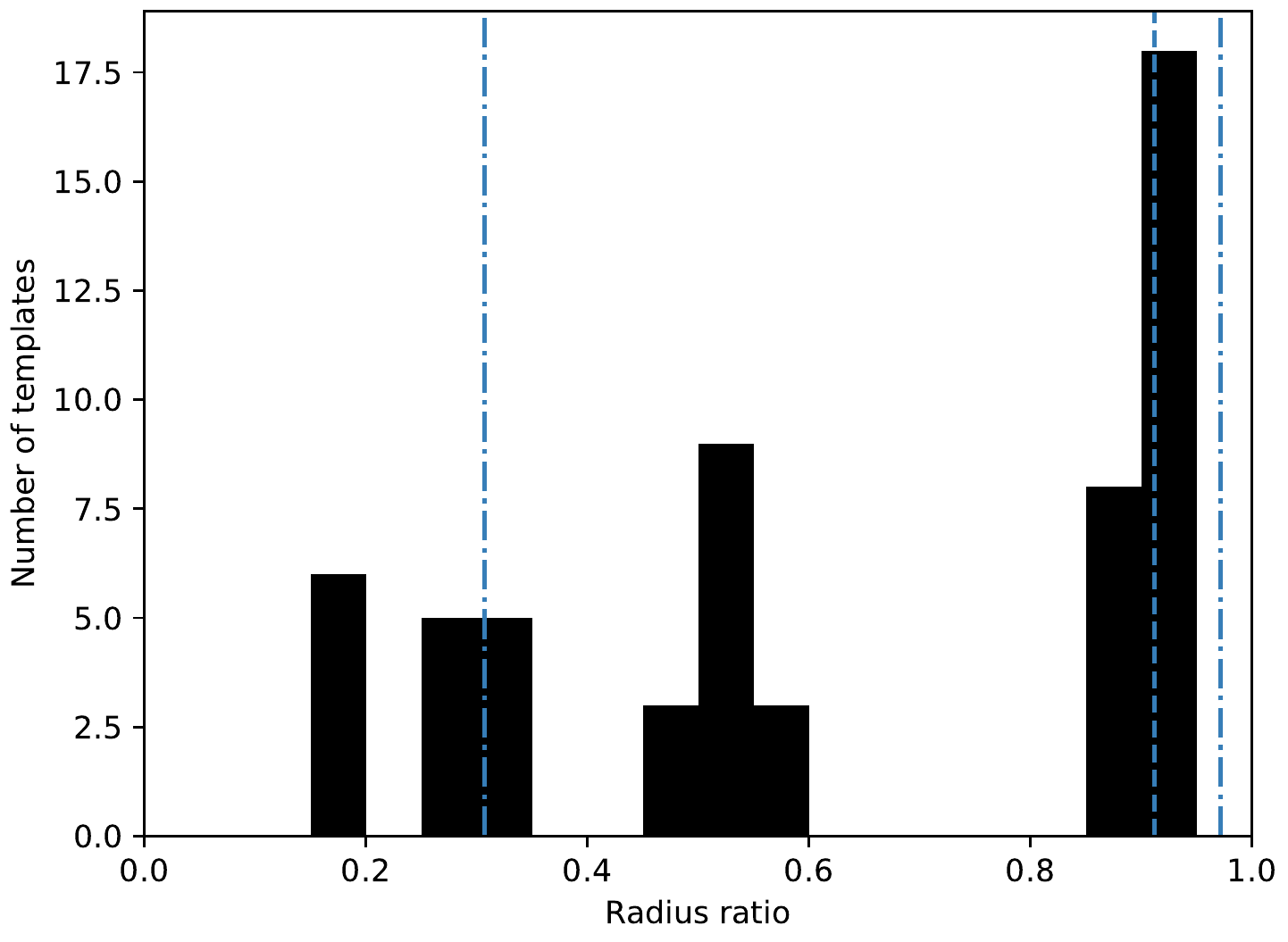}
    \caption{Distribution of radius ratios of the best fitting templates to the GHOST spectrum of \nigelsysname. The median radius ratio, $k_{\text{med}} = 0.91$ is marked as a vertical dashed line. The 68\% confidence interval, $[0.31,0.97]$, is shown as dash-dotted lines.}
    \label{fig:rr_dist}
\end{figure}

We use the mass ratio estimated in Section~\ref{sec:rv_meas}, $q=0.86\pm0.04$, to restrict the templates considered to have $0.72\leq q\leq1.00$ (the 99.99994\% confidence interval for $q$) and find a minimum \redchisq\ of 13.44.
This indicates that best fit template does not fit the data well, highlighting the challenges in the spectroscopic characterization of M-dwarf binaries, but we stress that the primary goal of this characterization is to provide initial estimates and priors for the eclipse and RV modeling in Section~\ref{sec:model}. We select all template fits with a \redchisq\ within 1 of this minimum \redchisq\ and find that the radius ratio, $k=\frac{R_B}{R_A}$, of the two components used to construct each template shows distinct clusterings for the best fitting templates, as seen in Figure~\ref{fig:rr_dist}.
We use these results of the spectroscopic characterization to perform four separate fits to the eclipse photometry and RV measurements in Section~\ref{sec:model}. We consider a model with a full range of radius ratios allowed ($0.15\leq k\leq1$, `Full $k$-range model') where this range is determined as the 99.99994\% confidence interval of radius ratio values for all the best fitting templates included in Figure~\ref{fig:rr_dist}. We also consider three models with restricted ranges for the radius ratio. The ranges of radius ratios for each model are determined by calculating the 99.99994\% confidence intervals of the radius ratios of the best fitting templates in each cluster shown in Figure~\ref{fig:rr_dist}. The clusters are broadly defined as: a `High $k$-range model' with $k>0.8$; a `Middle $k$-range model' with $0.4<k<0.7$; and a `Low $k$-range model' with $k<0.4$. The calculation of the confidence intervals then provides us with uniform priors for the radius ratio for each of the `High', `Middle' and `Low' models as: $0.86\leq k\leq1$, $0.48\leq k\leq0.56$ and $0.15\leq k\leq0.32$, respectively.
We show the first GHOST spectrum of \nigelsysname\ and the best-fitting template from the `High $k$-range model' for the ten spectral regions we fit in Figure~\ref{fig:spec_fit} in Appendix~\ref{sec:spec_fit}. We discuss the different models and the choice to show the best-fitting template from the `High $k$-range model' in Sections~\ref{sec:model} and Section~\ref{sec:discussion}.
We report the spectroscopic parameters (\teff, \logg\ and \feh) for all models in Table~\ref{tab:spec_params}. In some cases we find the upper or lower confidence interval values for some of the parameters are equivalent to the median value. Given the poor quality of the spectral fits we opt to impose minimum errors in these cases of 100\,K, 0.5\,log(cgs) and 0.5\,dex for the temperature, surface gravity and metallicity, respectively. These minimum values are determined from the step sizes in each parameter for the PHOENIX synthetic spectra.
\begin{table}[h]
\centering
\caption{Spectroscopic parameters for \nigelAname\ and \nigelBname\ for each of the models across different regions of the radius ratio parameter space. Uncertainties are the 68\% confidence intervals or minimum errors imposed as described in text.}\label{tab:spec_params}
\begin{tabular}{c|cccc}
Parameter (unit) & Full & Low & Middle & High \\ \hline
$T_{\rm eff,A}$ (K) & $3500^{+100}_{-200}$ & $3400^{+100}_{-100}$ & $3500^{+100}_{-100}$ & $3500^{+100}_{-200}$ \\
$\log g_A$ (log(cgs)) & $5.00^{+0.50}_{-0.50}$ & $4.50^{+0.50}_{-0.50}$ & $5.00^{+0.50}_{-0.50}$ & $5.50^{+0.50}_{-0.50}$ \\
$\mbox{[Fe/H]}_A$ (dex) & $0.0^{+0.5}_{-0.5}$ & $-0.5^{+0.5}_{-0.5}$ & $0.0^{+0.5}_{-0.5}$ & $0.0^{+0.5}_{-0.5}$ \\
$T_{\rm eff,B}$ (K) & $3300^{+100}_{-100}$ & $3300^{+100}_{-100}$ & $3300^{+100}_{-100}$ & $3300^{+100}_{-100}$ \\
$\log g_B$ (log(cgs)) & $5.50^{+0.50}_{-0.50}$ & $5.50^{+0.50}_{-0.50}$ & $5.50^{+0.50}_{-0.50}$ & $5.50^{+0.50}_{-0.50}$ \\
$\mbox{[Fe/H]}_B$ (dex) & $0.0^{+0.5}_{-0.5}$ & $0.0^{+0.5}_{-0.5}$ & $0.0^{+0.5}_{-0.5}$ & $0.0^{+0.5}_{-0.5}$ \\
\end{tabular}
\end{table}

\begin{figure}
    \centering
    \includegraphics[width=0.9\columnwidth]{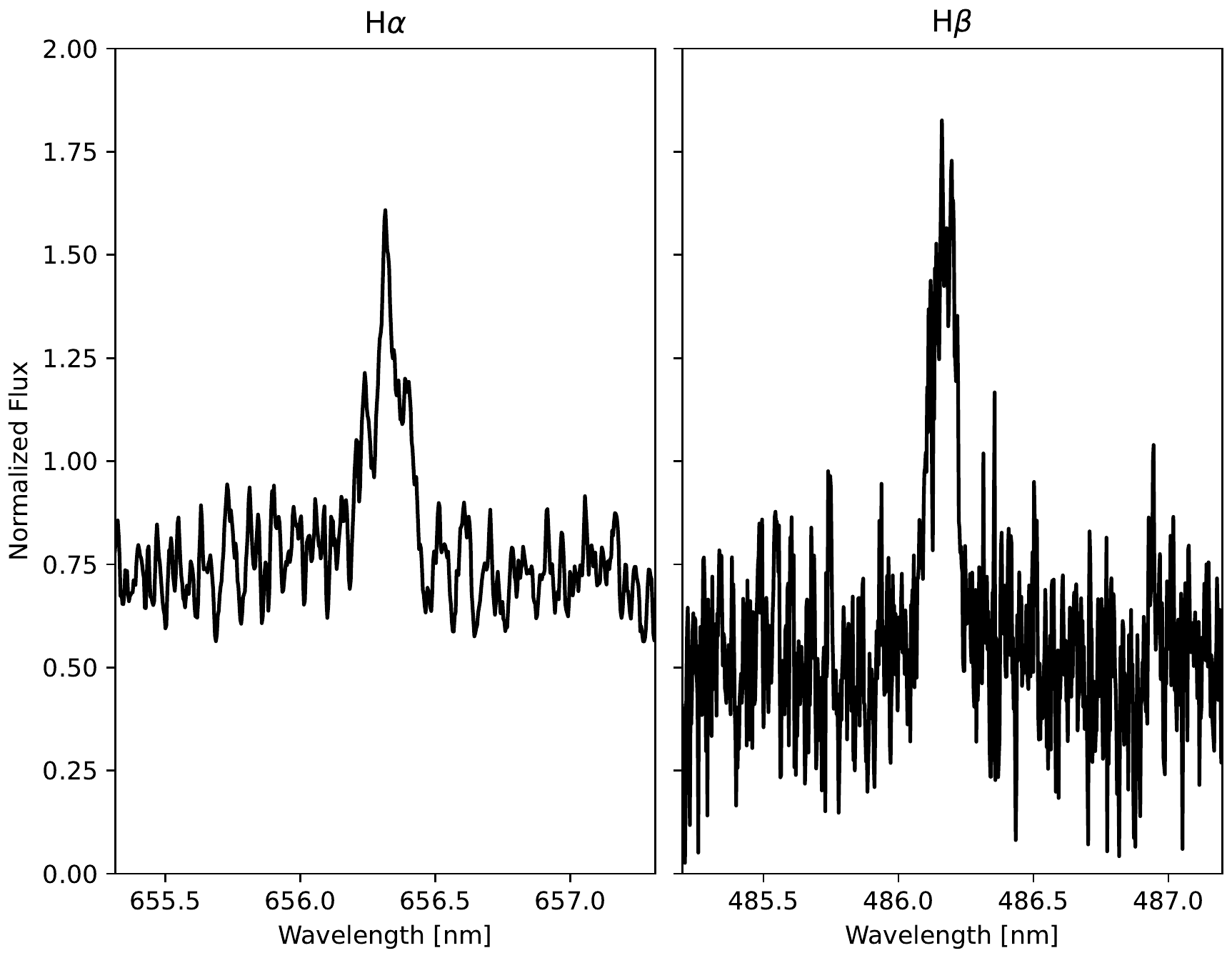}
    \caption{Normalized and stacked GHOST spectrum of \nigelsysname\ zoomed on the H$\alpha$ (left) and H$\beta$ (right) emission features.}
    \label{fig:halpha}
\end{figure}

Finally, Figure~\ref{fig:halpha} shows the weighted average of both GHOST spectra zoomed on the H$\alpha$ and H$\beta$ lines. We detect emission in these lines which is often an indicator of stellar activity \citep[e.g.][]{Berger2006LowMassMagnets,Scholz2007PreMainActivity,Stassun2012LowMassActivity}. We are unable to determine whether the emission is from one or both stellar components.

\section{System Modeling}\label{sec:model}
We use the \allesfitter\ package \citep{Guenther2019allesfittercode,Guenther2021allesfitterpaper} to simultaneously fit the photometric and spectroscopic data and determine the system parameters for each of the 4 models described in Section~\ref{sec:spec_char}.

We use the stellar atmospheric parameters in Table~\ref{tab:spec_params} and the \texttt{PyLDTk} package \citep{Parviainen2015LDTk}, that utilizes Phoenix V2 models \citep{Husser2013PhoenixV2} and transmission curves from the Spanish Virtual Observatory (SVO) Filter Service \citep{Rodrigo2012SVO,Rodrigo2020SVO}, to estimate initial guesses for the limb darkening coefficients for both stars in all filters for each of the models considered. We adopt a quadratic limb darkening law in all cases.
We couple the limb darkening coefficients for all \tess\ sectors to reduce the number of free parameters.

For each model, we estimate the surface brightness ratios of the two stars in all photometric filters as follows. We convolve the PHOENIX spectra that were used to construct the SB2 templates described in Section~\ref{sec:spec_char} with the transmission curves for each photometric filter. We numerically integrate under these convolved spectra using Simpson's rule and calculate the ratio of these values for each of the best fitting SB2 templates. We then set the initial guess for each surface brightness ratio as the median value, $\text{SB}_{\text{med}}$. We use a normal distribution for the prior, centered on $\text{SB}_{\text{med}}$ with a standard deviation $\sigma = \text{max}[\text{SB}_{\text{med}} - \text{SB}_{\text{LL}}, \text{SB}_{\text{UL}} - \text{SB}_{\text{med}}]$, where $\text{SB}_{\text{LL}}$ and $\text{SB}_{\text{UL}}$ are the lower and upper 68\% confidence intervals of the surface brightness estimates.
We couple these parameters for all \tess\ sectors to reduce the number of free parameters.

We implement the Mat\'{e}rn 3/2 Gaussian Process (GP) kernel to model any photometric variability as this kernel is versatile in its ability to model both long and short-term trends \citep{Guenther2021allesfitterpaper}.
The GP hyperparameters for each instrument are estimated from out-of-transit data and then fixed for the full fits to reduce the number of free parameters. To test the impact of including a GP in our modeling, we run a fit without the GP for any instrument. The posteriors of the fit without the GP are erratic and the logarithmic Bayes factors comparing the GP fits with the non-GP fit are $\Delta \ln Z \gtrapprox 70000$, indicating that the GP fits are overwhelmingly preferred. Therefore, we adopt the parameters determined from the fits including the GP kernel to model any variability in the light curves.

Given the long exposure times used for \tess\ Sector 10 ($t_{\text{exp}} = 1800$\,s), \tess\ Sector 37 ($t_{\text{exp}} = 600$\,s) and the SAAO observations taken on 2023 February 28 with the $g^\prime$ filter ($t_{\text{exp}} = 600$\,s), we use finer sampling grids for the light curve models for each of these photometric datasets. We use grids 10 and 5 times finer for the 1800\,s and 600\,s exposure time data, respectively (see \citet{Guenther2021allesfitterpaper} for details).
Due to the large pixel scale of \tess\ (21\,arcsec\,per\,pixel), the \tess\ data are fit with a dilution factor to account for any contributions from nearby stars.

Despite fixing multiple parameters, due to the large number of instruments used there remains 55 free parameters. We attempted using a Markov chain Monte Carlo (MCMC) approach however the fits did not converge after 200,000 steps. Therefore, we employ a nested sampling approach rather than MCMC to ensure the fit converges in a reasonable timescale \citep{Guenther2021allesfitterpaper}.
We use the default Nested Sampling settings for \texttt{dynesty} recommended in \citet{Guenther2021allesfitterpaper}, that is: algorithm, $dynamic$; number of live points, $n_{\text{live}}=500$; prior bound method, $single$; live point update method, $rwalk$; and tolerance of convergence criterion, $n_{\text{tol}}=0.01$. We refer the reader to \citet{Speagle2020dynesty} and \citet{Guenther2021allesfitterpaper} for more detail.

\texttt{dynesty} provides a measure of the logarithmic Bayesian evidence for each model that we can use to compare how well each model fits the data. We compare two different models by calculating the logarithmic Bayes factor $\Delta \ln Z_{2,1} = \ln Z_2 - \ln Z_1$ where $\ln Z_2$ and $\ln Z_1$ are the logarithmic Bayesian evidences for given models. We find the logarithmic Bayesian evidence for the low $k$-range model to be $\ln Z_{\text{low}} = 38682.1 \pm 0.3$. This is significantly smaller ($\Delta \ln Z_{n,\text{low}} > 4000$) than the evidences calculated for each of the other 3 models ($n = \text{full, middle, high}$) and therefore we discard this model from further consideration.

\begin{figure*}
    \centering
    \includegraphics[width=0.99\columnwidth]{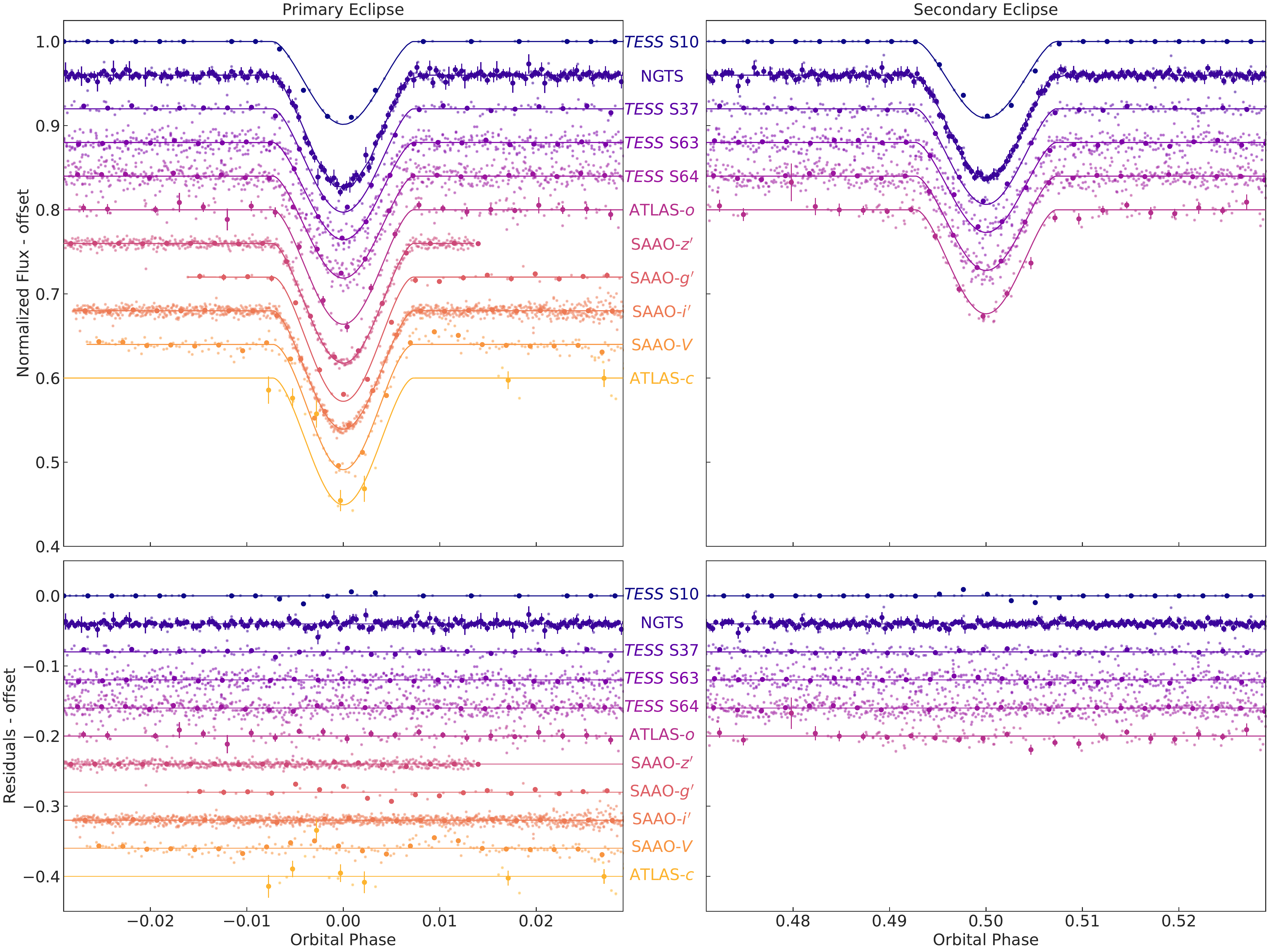}
    \caption{Discovery and follow-up photometry obtained for \nigelsysname. (Top left) Phase-folded photometry for the primary eclipse in each instrument, offset for clarity. Binned data in 2\,min bins for NGTS and 15\,min bins for all other instruments is shown as solid circles with raw data shown as lower opacity points. The median best fit \allesfitter\ model for the High $k$-range model is shown as a solid line for each photometric dataset. (Top right) Phase-folded photometry for the secondary eclipse in each instrument it was observed, offset for clarity. (Bottom row) Residuals for the primary and secondary eclipses in each instrument, offset for clarity. The photometric data are provided in the supplementary material.}
    \label{fig:alles_photom}
\end{figure*}

\begin{figure*}
    \centering
    \includegraphics[width=0.95\columnwidth]{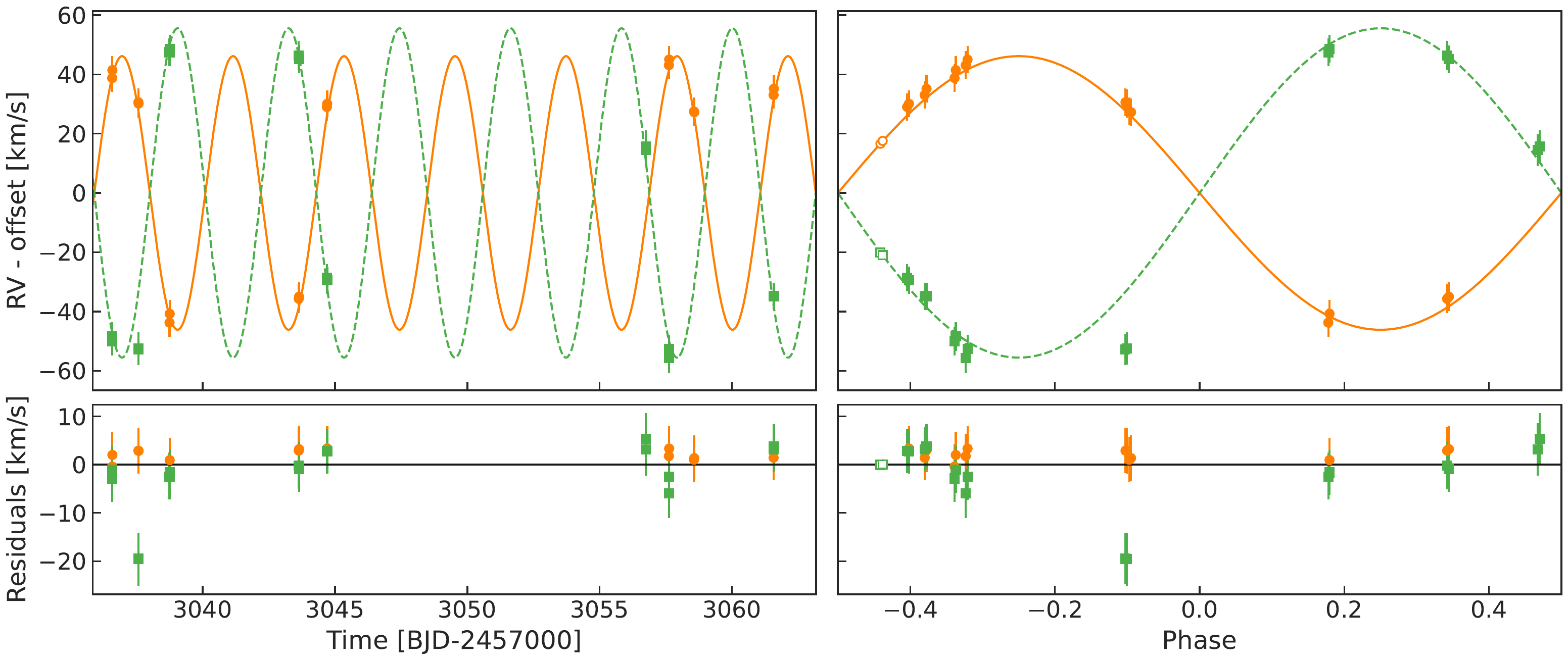}
    \caption{(Top left) The NIRPS radial velocity data against time. (Top right) The phase-folded NIRPS and GHOST radial velocity data. The RVs for the primary star are shown as orange circles (NIRPS: solid circles; GHOST: open circles) and the median \allesfitter\ model for the High $k$-range model is shown as a solid orange line. The RVs for the secondary star are shown as green squares (NIRPS: solid squares; GHOST: open squares) and the median \allesfitter\ model is shown as a dashed green line. (Bottom row) Residuals for the RVs shown in both panels.}
    \label{fig:alles_rv}
\end{figure*}

We report the key fitted and derived parameters for \nigelsysname\ for the three remaining models in Table~\ref{tab:sys_params}. The derived parameters are calculated using the series of equations presented in \citet{Hilditch2001CloseBinaryStars}, to which we refer the reader for further details. We also include the logarithmic Bayesian evidence for each model calculated by \texttt{dynesty}.
\begin{table}
    \centering
    \caption{\nigelsysname\ system parameters obtained with \allesfitter\ for the Full $k$-range, Middle $k$-range and High $k$-range models.}
    \label{tab:sys_params}
    \begin{tabular}{cccc}
    \hline
    Parameter (unit) & Full & Middle & High \\ 
    \hline
    \multicolumn{4}{c}{\textit{Fitted parameters}} \\
    $T_{0;B}$ (BJD) & \Bepochfull & \Bepochmid & \Bepochhigh \\
    $P$ (d) & \Bperiodfull & \Bperiodmid & \Bperiodhigh \\ 
    $R_B/R_A$ & \Brrfull & \Brrmid & \Brrhigh \\
    $(R_A+R_B)/a$ & \Brsumafull & \Brsumamid & \Brsumahigh \\
    cos($i$) & \Bcosifull & \Bcosimid & \Bcosihigh \\
    $K_A$ (\kms) & \KAfull & \KAmid & \KAhigh \\
    Mass ratio, $q$ & \Bqfull & \Bqmid & \Bqhigh \\
    \multicolumn{4}{c}{\textit{Derived parameters}} \\
    $K_B$ (\kms) & \KBfull & \KBmid & \KBhigh \\
    $i$ (deg) & \incdegfull & \incdegmid & \incdeghigh \\
    $a$ (au) & \semimajaufull & \semimajaumid & \semimajauhigh \\
    $M_A$ (\msun) & \MAfull & \MAmid & \MAhigh \\
    $M_B$ (\msun) & \MBfull & \MBmid & \MBhigh \\
    $R_A$ (\rsun) & \RAfull & \RAmid & \RAhigh \\
    $R_B$ (\rsun) & \RBfull & \RBmid & \RBhigh \\
    $\log g_A$ ($\log$(cgs)) & \loggAfull & \loggAmid & \loggAhigh \\
    $\log g_B$ ($\log$(cgs)) & \loggBfull & \loggBmid & \loggBhigh \\
    \multicolumn{4}{c}{\textit{Bayesian Evidence}} \\
    $\ln Z$ & \bayesfull & \bayesmid & \bayeshigh \\
    \hline
    \end{tabular}
\end{table}
We see that, other than the radius ratio, the fitted parameters are broadly consistent across the three models. The radius ratio for the Full and Middle $k$-range models are consistent within the uncertainties while the radius ratio for the High $k$-range model is much larger. We see that the Bayesian evidences suggest preference towards the Full and Middle $k$-range models with $\Delta \ln Z_{\text{Full},\text{High}}$ and $\Delta \ln Z_{\text{Middle},\text{High}}$ both greater than 3, indicating strong evidence in favor of these models compared with the High $k$-range model, with a slight preference toward the Full $k$-range \citep{KassRaftery1995Bayes,Jeffreys1998theoryProb,Trotta2008bayes}. However, the differences in the radius ratios propagate into the determinations of the stellar radii ($R_A$ and $R_B$) and, as later discussed in Section~\ref{sec:mmcomp}, lead to significant differences between the derived radii for the Full and Middle $k$-range models, and the M-dwarf mass-radius relations presented in Section~\ref{sec:mmcomp}. We therefore opt to show the best fit model from \allesfitter\ for the High $k$-range model in Figures~\ref{fig:alles_photom}~and~\ref{fig:alles_rv} that show the photometric data and radial velocities, respectively.

\begin{figure}
    \centering
    \includegraphics[width=0.9\columnwidth]{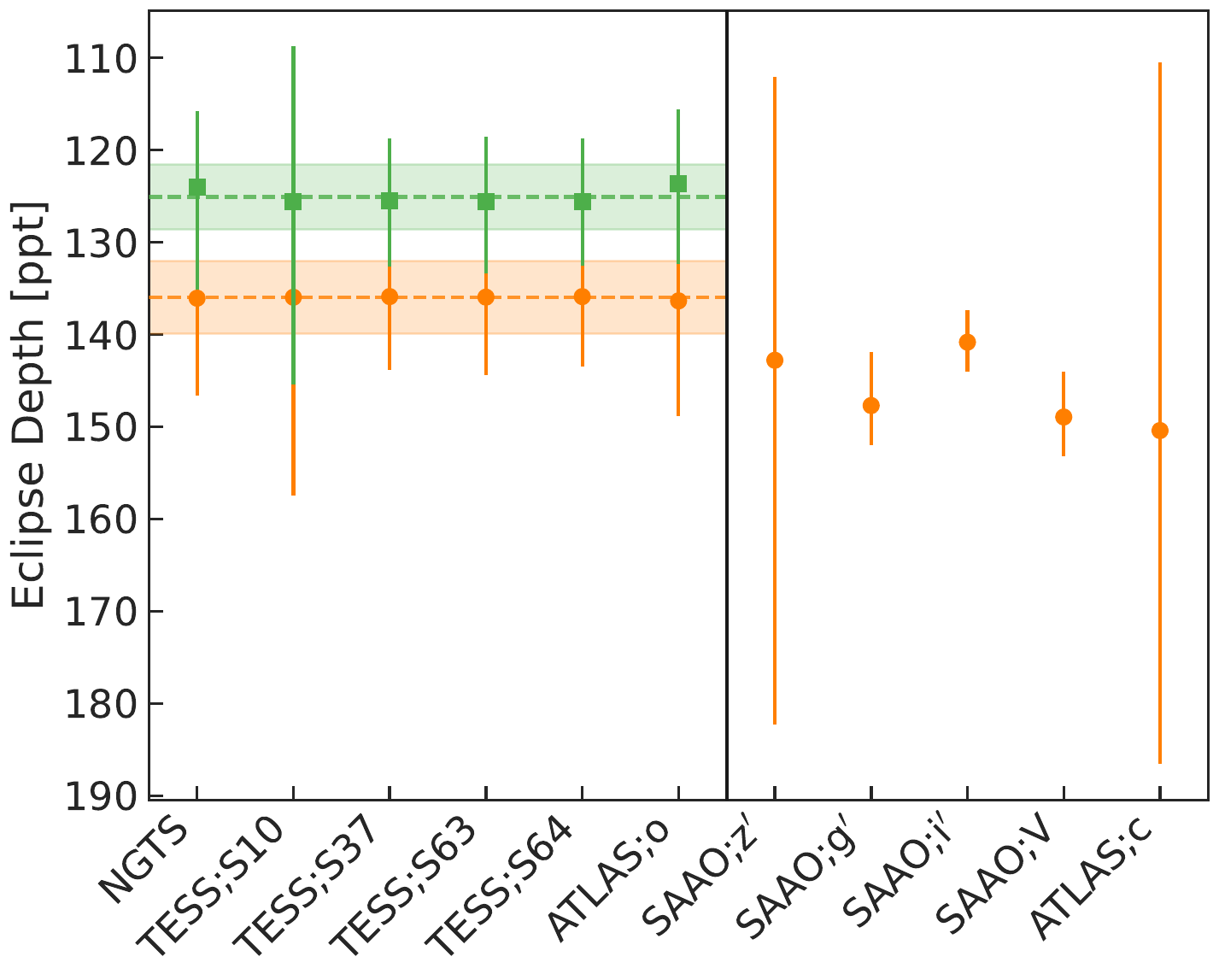}
    \caption{Eclipse depths for \nigelsysname\ derived for the primary (orange circles) and secondary (green squares) eclipses in parts per thousand. The uncertainties are the 99.7\% ($\mysim3\sigma$) percentiles of the derived posteriors. (Left panel) Eclipse depths for instruments where we detect both the primary and secondary eclipses. The weighted mean for both eclipses is shown as a dashed line with shaded regions indicating the weighted uncertainty. (Right panel) Primary eclipse depths for the SAAO and ATLAS-$c$ observations.}
    \label{fig:eclipse_depths}
\end{figure}

We derive the eclipse depths for both the primary and secondary eclipses across each instrument. This is to test two phenomena: whether a clear depth difference exists between the primary and secondary eclipses, a commonly used test for binarity in exoplanet transit searches \citep{ODonovan2006TrES_BEBReject,Howell2011SpeckleKepler,LilloBox2014ComprehensiveFPRejection,Ciardi2015StellarMultiplicity,Lester2021SpeckleTESS}; and whether the systems exhibit eclipses of varying depths when viewed in multiple bandpasses, which is an indicator that a system is composed of two stars of differing colors \citep{Rosenblatt1971MulticolourTheory,Drake2003ColorPhotom_FPs,Tingley2004ColorPhotom_FPs,Parviainen2019ColorPhotom_FPs}.
These eclipse depth measurements are shown in Figure~\ref{fig:eclipse_depths}.
We find the observations of the primary eclipse using SAAO with multiple filters are also broadly consistent within the uncertainties, thus we are unable to detect any variation in the eclipse depths that would have been indicative of the system being composed of two stars of differing colors.
We identify a systematic depth difference between the primary and secondary eclipse depths across all bandpasses where we observe both eclipses, however we note the difference is not significant in any one dataset.
We measure the depth difference to be less than 15\,ppt in all bandpasses, while we measure the standard deviation of the flux of each light curve ($\sigma_{\text{LC}}$) to be greater than 15\,ppt.
\citetalias{Bryant2023OccurrenceGiantsMDwarfs} utilized only the \tess\ Sector 10 light curve in which the depths are fully consistent within the reported uncertainties, therefore the candidate successfully passed the odd/even check described in their work. Likewise, this depth difference was not identified during the vetting and analysis of the data described in \citet{OBrien2024PHNGTSPaper1}.
For instruments where we have detections of both the primary and secondary eclipses, we estimate the weighted mean of the primary and secondary eclipses to be $\delta_p = 135.97 \pm 3.96$\,ppt and $\delta_s = 125.08 \pm 3.52$\,ppt, respectively.
We conclude that despite the systematic difference in the primary and secondary eclipse depths, the difference is marginal relative to the scatter of the light curves so we are unable to have definitively identified this system as a binary from the odd/even check alone.

\section{Flare Environment}\label{sec:flares}
Stellar flares are the consequence of the reconnection of magnetic field lines, resulting in large, rapid brightening events across the electromagnetic spectrum \citep[e.g.][and references therein]{Shibata1999UnifiedFlareModel,Benz2010FlareReview}.
It has been determined that M-dwarf flare energies can span a wide range \citep[$10^{26} - 10^{36}$\,erg;][]{Lacy1976UVCetiFlares,Kowalski2010MegaflareYZCMi,Davenport2016KeplerFlares,Gunther2020TESSFlares}, with these events observed often as these low-mass stars remain magnetically active for much longer than FGK-type stars \citep{West2008AgeActivityRelation,Hilton2010MDwarfFlaresSDSS,Pineda2013SDSSMDwarfActivity}. Flares can present a challenge to the habitability of exoplanets (and any potential exomoons) that reside close to M-dwarfs. The habitable zone of these host stars are much closer (10-100 times) than that of the Sun, greatly increasing the effect of a flare on potentially habitable planetary bodies. Magnetic activity and flaring can cause issues such as atmospheric erosion \citep{Lammer2007CMEHabitability}, runaway greenhouse effects and hydrodynamic escape of the atmospheres \citep{Luger2015RunawayGreenhouse,Shields2016HabitabilityMDwarfs}. Therefore the characterization of the flare activity of M-dwarfs, in particular those in systems on short-period orbits, is of particular importance in this context.

To assess the flare environment of the \nigelsysname\ system, we search all available high-cadence, photometric data (NGTS, SAAO, \tess) for candidate flare events using a method similar to that described in \citet{Jackman2018GStarSuperflares,Jackman2019PMSMFlare,Jackman2020OrionFlares,Jackman2021NGTSFlares}. Prior to applying the flare search algorithm, we subtract the transit model provided by \allesfitter\ from all datasets. We look for potential flares within each night of the NGTS and SAAO data by searching for regions where there are at least three consecutive points above six median absolute deviations (\mad) from the median of the night. We also search for any night where the median flux for the night is at least five \mad\ above the median of the whole light curve for the NGTS data. This is to test for long duration, high-amplitude flares that may dominate an entire night of observation. Given the SAAO data are taken in different filters and we have only four nights of observations, we do not compare between separate nights. Similarly, for each \tess\ sector, we search for at least three consecutive points above six \mad\ from the median flux of the entire sector. We detect no candidate flare events in these datasets. Visual inspection of each individual night/dataset also revealed no flare events in the high-cadence photometry.

\begin{figure*}
    \centering
    \includegraphics[width=0.95\columnwidth]{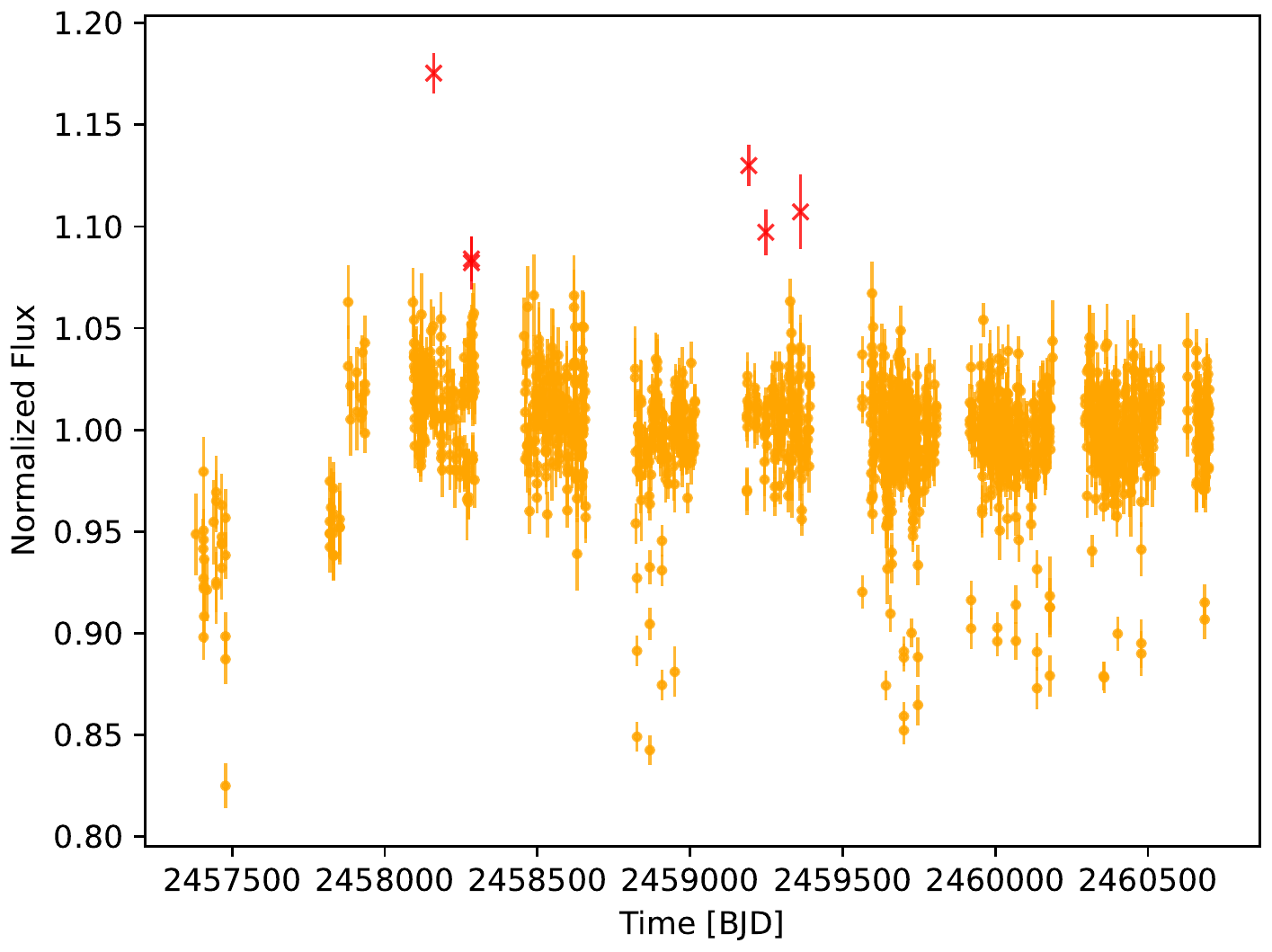}
    \caption{ATLAS $o$-band data for \nigelsysname\ with candidate flare events highlighted as red crosses.}
    \label{fig:atlas_flare}
\end{figure*}

We also search the ATLAS data (Section~\ref{sec:atlas}) for possible flares using a flare identification criteria similar to those described in \citet{Kowalski2009MDwarfSDSS}, \citet{Hawley2014KeplerMDwarfFlares} and \citet{RodriguezMartinez2020ASASSNFlares}.
We identify candidate flares as events that satisfy
\begin{equation}
    \frac{f - f_{\text{median}}}{\sigma} \geq 3, \label{eq:atlas_flare}
\end{equation}
where $f$ is the normalized flux following the relative photometry described in Section~\ref{sec:atlas}, $f_{\text{median}}$ is the median flux of the light curve and $\sigma$ is the standard deviation of the light curve. The median is chosen as it is less sensitive to outliers.
We elect to use a $3\sigma$ cut, following \citet{Kowalski2009MDwarfSDSS} as we find that the lower cuts of $2.5\sigma$ \citep{Hawley2014KeplerMDwarfFlares} and $2\sigma$ \citep{RodriguezMartinez2020ASASSNFlares} flag events that broadly lie within the scatter of the light curve.
The ATLAS-$o$ light curve is shown in Figure~\ref{fig:atlas_flare} with candidate flares that meet the criteria in Equation~\ref{eq:atlas_flare} highlighted in red.
We do not identify flares in the ATLAS-$c$ data so exclude this data from Figure~\ref{fig:atlas_flare} for clarity.
We identify 5 candidate flare events that warrant future investigation. Combined with the H$\alpha$/H$\beta$ emission detected in the GHOST spectra (Section~\ref{sec:ghost}), this may be indicative of some level of stellar activity from at least one of the components in this system \citep[e.g.][]{Berger2006LowMassMagnets,Scholz2007PreMainActivity,Stassun2012LowMassActivity,Medina2020MDwarfFlares}.
We re-iterate however that we do not detect flares in any of the high-cadence photometric data, thus the activity of this system is unclear and requires further monitoring.

\section{Discussion}\label{sec:discussion}
\subsection{Comparison with M+M binary population}\label{sec:mmcomp}
\begin{figure*}
    \centering
    \gridline{\fig{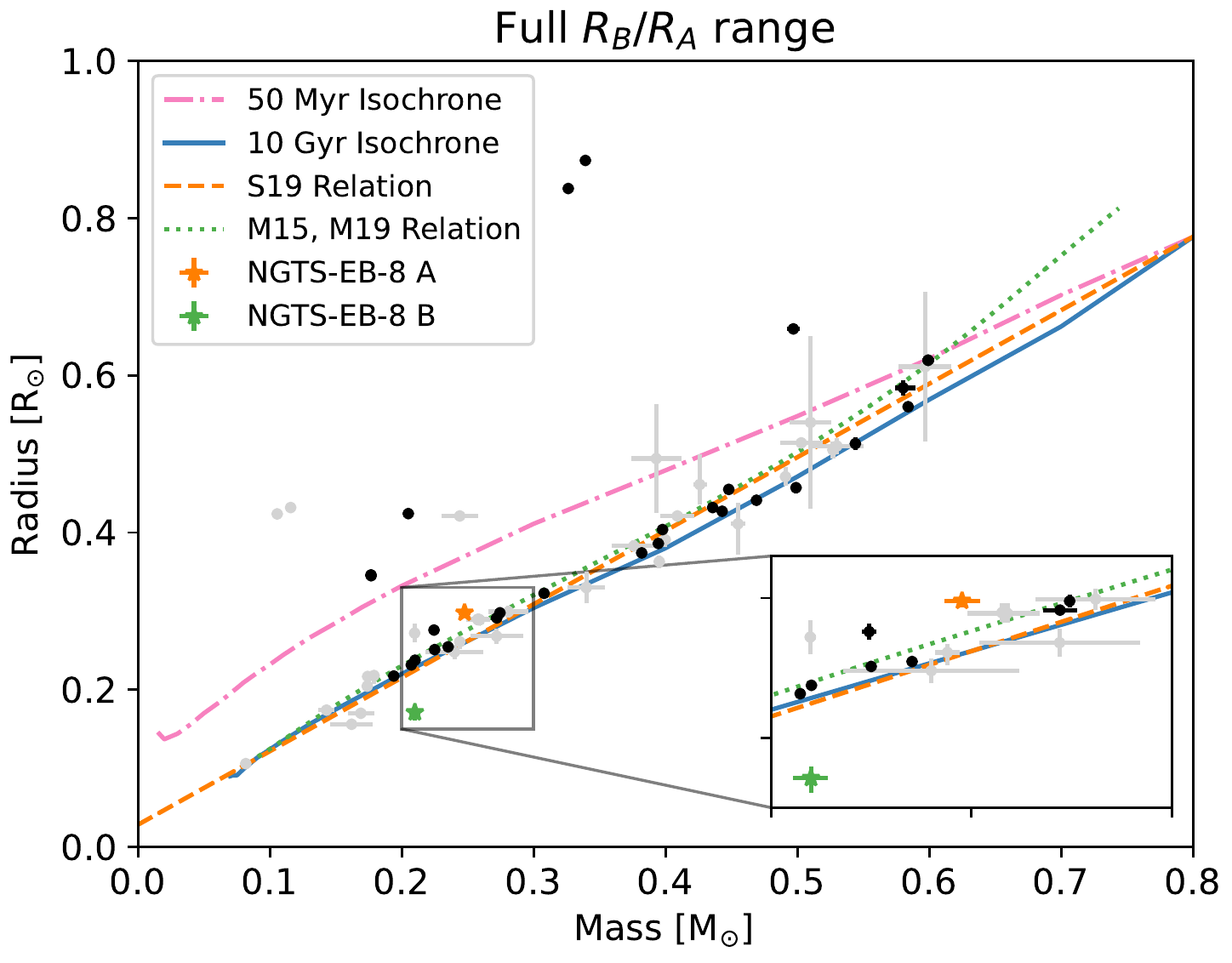}{0.49\textwidth}{}
          \fig{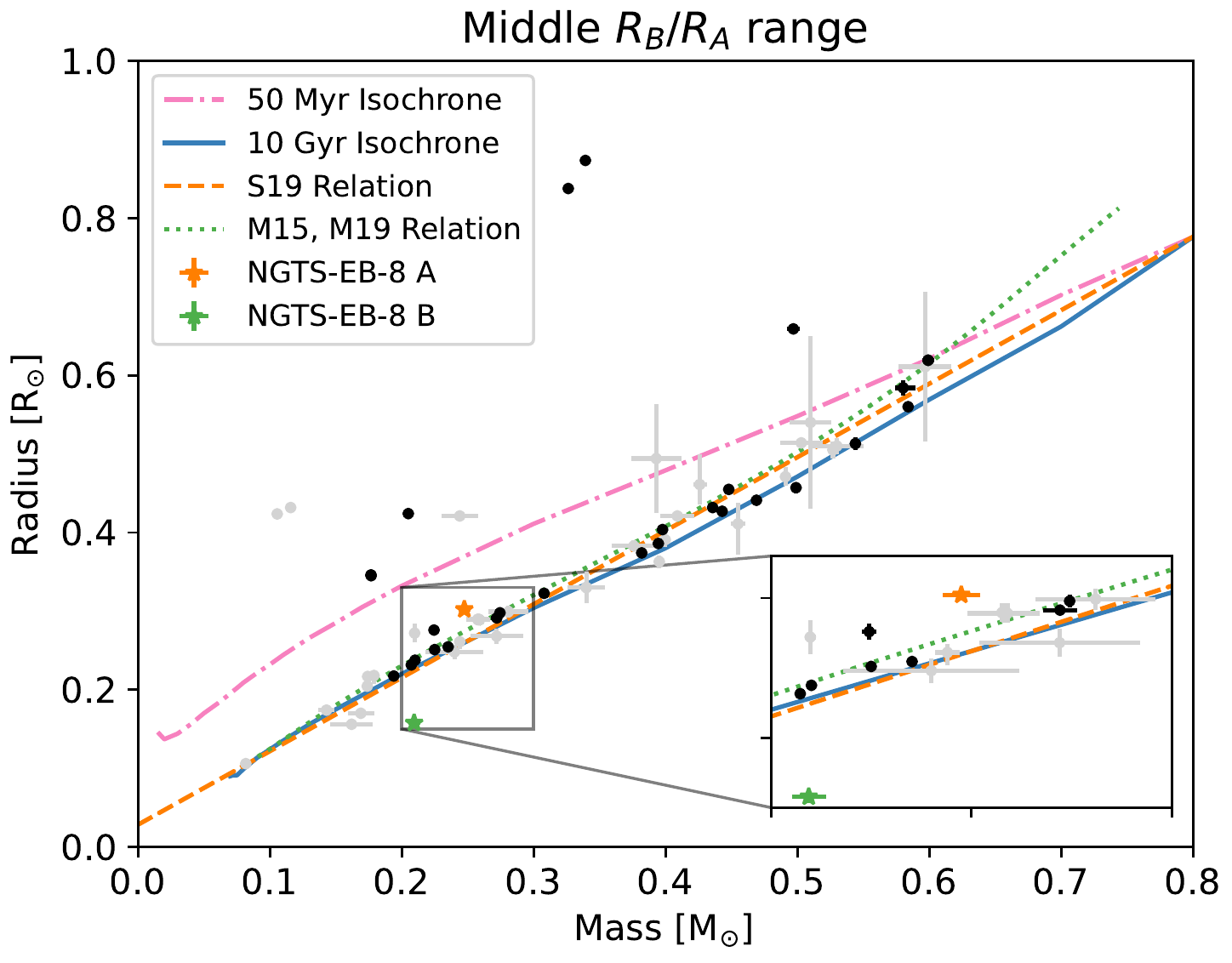}{0.49\textwidth}{}}
          \vspace{-1cm}
          \gridline{\fig{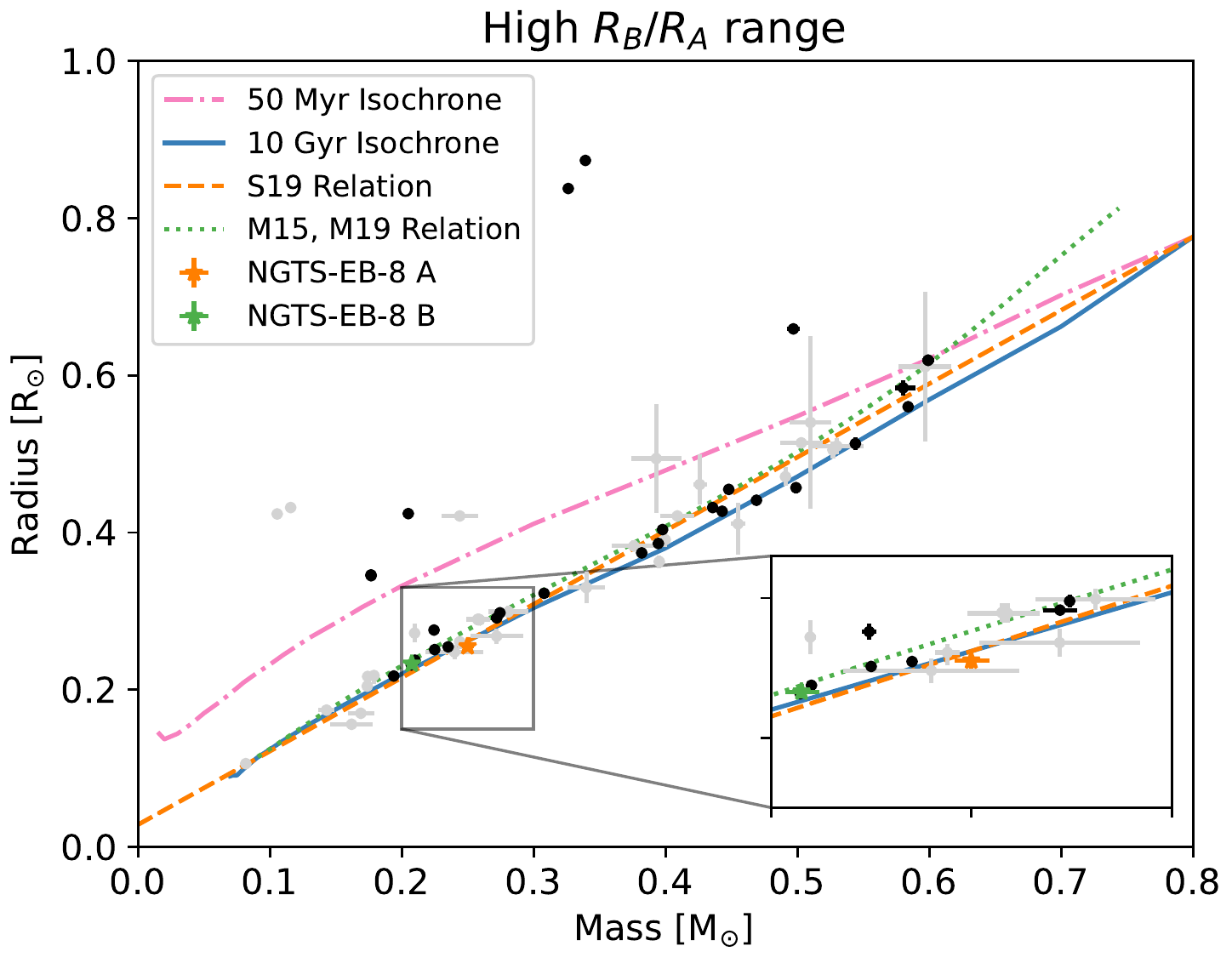}{0.49\textwidth}{}}
          
          \caption{Mass-radius diagrams of low-mass stars in double-lined eclipsing binary systems composed of two M-dwarfs with the position of the components of \nigelsysname\ shown for the different $k$-range models in each panel. In all panels, black points show stars with both mass and precision measured to better than 2\%. Light grey points show all other stars found in the literature search (see Table~\ref{tab:mm_table}). \nigelAname\ is shown as an orange star. \nigelBname\ is shown as a green star. The 50\,Myr and 10\,Gyr isochrones from \cite{Baraffe2015Isochrones} are plotted as dash-dotted pink and solid blue lines, respectively. The relation from \citet{Schweitzer2019CARMENESTargetStars} is shown as a dashed orange line. The relation derived from \citet{Mann2015ConstrainMDwarfs} and \citet{Mann2019ConstrainMdwarfII} is shown as a dotted green line.
          (Top left panel) Positions of \nigelsysname\ shown as determined by the Full $k$-range model. (Top right panel) Positions of \nigelsysname\ shown as determined by the Middle $k$-range model. (Bottom panel) Positions of \nigelsysname\ shown as determined by the High $k$-range model.}\label{fig:binarypop}
\end{figure*}

We perform a literature search for other double-lined eclipsing binary systems composed of two M-dwarfs. These systems are listed in  Table~\ref{tab:mm_table}.
We plot these systems, the positions of \nigelsysname\ determined from each of the models with different $k$-ranges, and a variety of mass-radius relations from literature in Figure~\ref{fig:binarypop}. We include the 50\,Myr and 10\,Gyr isochrone models from \citet{Baraffe2015Isochrones}, the mass-radius relation from Equation~7 in \citet{Schweitzer2019CARMENESTargetStars} and the mass-radius relation derived by combining the radius-magnitude and mass-magnitude relations presented in \citet{Mann2015ConstrainMDwarfs} and \citet{Mann2019ConstrainMdwarfII}. These diagrams demonstrate the radius inflation problem of low-mass stars as many of the stars have observed radii larger than those predicted by the theoretical models.
We compare the positions of \nigelAname\ and \nigelBname\ to each of the \citet{Baraffe2015Isochrones} 10\,Gyr isochrone, the \citet{Schweitzer2019CARMENESTargetStars} relation and the \citet{Mann2015ConstrainMDwarfs,Mann2019ConstrainMdwarfII} relation in Table~\ref{tab:inflation}.
\begin{table}
\centering
\caption{Positions of the components of \nigelsysname\ relative to the various mass-radius relations for the parameters derived from the Full, Middle, and High $k$-range models. The top number in each cell is the sigma-scaled shortest distance to each mass-radius relation. The bottom number in each cell is the discrepancy in the radius expressed as a percentage (assuming the best-fit mass) for each relation. \citetalias{Baraffe2015Isochrones} - \citet{Baraffe2015Isochrones} 10 Gyr isochrone; \citetalias{Schweitzer2019CARMENESTargetStars} - \citet{Schweitzer2019CARMENESTargetStars} relation; \citetalias{Mann2015ConstrainMDwarfs}, \citetalias{Mann2019ConstrainMdwarfII} - Combined \citet{Mann2015ConstrainMDwarfs} and \citet{Mann2019ConstrainMdwarfII} relation.}\label{tab:inflation}
\begin{tabular}{c|cc||cc||cc}
& \multicolumn{6}{c}{$k$-range Model} \\ \hline
                         & \multicolumn{2}{c||}{Full} & \multicolumn{2}{c||}{Middle} & \multicolumn{2}{c}{High} \\ \cline{2-7} 
Relation                 & \multicolumn{1}{c|}{Star A} & Star B & \multicolumn{1}{c|}{Star A} & Star B & \multicolumn{1}{c|}{Star A} & Star B \\ \hline
\multirow{2}{*}{\citetalias{Baraffe2015Isochrones} 10\,Gyr} 
  & \multicolumn{1}{c|}{6.63$\sigma$} & 5.17$\sigma$ & \multicolumn{1}{c|}{9.68$\sigma$} & 11.94$\sigma$ & \multicolumn{1}{c|}{1.05$\sigma$} & 1.09$\sigma$ \\
  & \multicolumn{1}{c|}{$114.6^{+1.7}_{-1.9}$\%} & $74.9^{+3.7}_{-4.6}$\% & \multicolumn{1}{c|}{$116.3^{+0.8}_{-0.9}$\%} & $69.2^{+2.0}_{-1.9}$\% & \multicolumn{1}{c|}{$97.4^{+1.7}_{-2.1}$\%} & $102.9^{+2.6}_{-2.2}$\% \\ \hline
\multirow{2}{*}{\citetalias{Schweitzer2019CARMENESTargetStars}} 
  & \multicolumn{1}{c|}{6.38$\sigma$} & 4.80$\sigma$ & \multicolumn{1}{c|}{8.97$\sigma$} & 10.99$\sigma$ & \multicolumn{1}{c|}{0.97$\sigma$} & 1.53$\sigma$ \\
  & \multicolumn{1}{c|}{$114.7^{+1.7}_{-1.9}$\%} & $76.2^{+3.8}_{-4.6}$\% & \multicolumn{1}{c|}{$116.4^{+0.8}_{-0.9}$\%} & $70.4^{+2.1}_{-2.0}$\% & \multicolumn{1}{c|}{$97.4^{+1.7}_{-2.1}$\%} & $104.8^{+2.6}_{-2.2}$\% \\ \hline
\multirow{2}{*}{\citetalias{Mann2015ConstrainMDwarfs}, \citetalias{Mann2019ConstrainMdwarfII}} 
  & \multicolumn{1}{c|}{4.04$\sigma$} & 6.17$\sigma$ & \multicolumn{1}{c|}{6.16$\sigma$} & 13.24$\sigma$ & \multicolumn{1}{c|}{3.09$\sigma$} & 0.71$\sigma$ \\
  & \multicolumn{1}{c|}{$108.6^{+1.6}_{-1.8}$\%} & $71.4^{+3.5}_{-4.3}$\% & \multicolumn{1}{c|}{$110.3^{+0.8}_{-0.8}$\%} & $65.9^{+1.9}_{-1.8}$\% & \multicolumn{1}{c|}{$92.3^{+1.6}_{-2.0}$\%} & $98.1^{+2.4}_{-2.1}$\%       
\end{tabular}
\end{table}
We find strong agreement between the parameters determined by the High $k$-range model for both stars and the \citet{Baraffe2015Isochrones} and \citet{Schweitzer2019CARMENESTargetStars} relations. We note that both stars lie below the relation derived from \citet{Mann2015ConstrainMDwarfs} and \citet{Mann2019ConstrainMdwarfII}, with \nigelAname\ at $3.09\sigma$ below the relation, however this is also the case for multiple other stars from literature across the low-mass regime as can be seen in Figure~\ref{fig:binarypop}.
We see that for the Full and Middle $k$-range models, \nigelAname\ shows $\mysim10-15\%$ inflation compared to each relation and deviates by greater than $4\sigma$ in all comparisons (with the sigma-scaled distance to the \citet{Baraffe2015Isochrones} 10\,Gyr isochrone being as high as $9.68\sigma$).
Given the smaller radius ratios and therefore comparatively smaller $R_B$ values derived by these models, we see that \nigelBname\ apparently lies significantly below the radius values predicted by all three relations. However, given that we know of no known mechanisms for radius deflation of low-mass stars, as well as the better agreement between the derived parameters and the mass-radius relations, we believe the parameters determined by the High $k$-range model are likely to be more accurate. Hence we opt to display the best fitting models from this fit in Figures~\ref{fig:alles_photom},~\ref{fig:alles_rv}~and~\ref{fig:spec_fit}.
The position of both stars compared to the \citet{Mann2015ConstrainMDwarfs}, \citet{Mann2019ConstrainMdwarfII} relations as well as the sensitivity of the best-fitting parameters and Bayesian evidences to the priors chosen suggest that further spectroscopy is required to more accurately determine the stellar atmospheric parameters and further measurements are required to increase the precision of the mass and radius measurements of this system to draw any further conclusions in the context of the radius inflation of low-mass stars.

\subsection{Comparison with Planet Candidates}
\begin{figure}
    \centering
    \includegraphics[width=0.9\columnwidth]{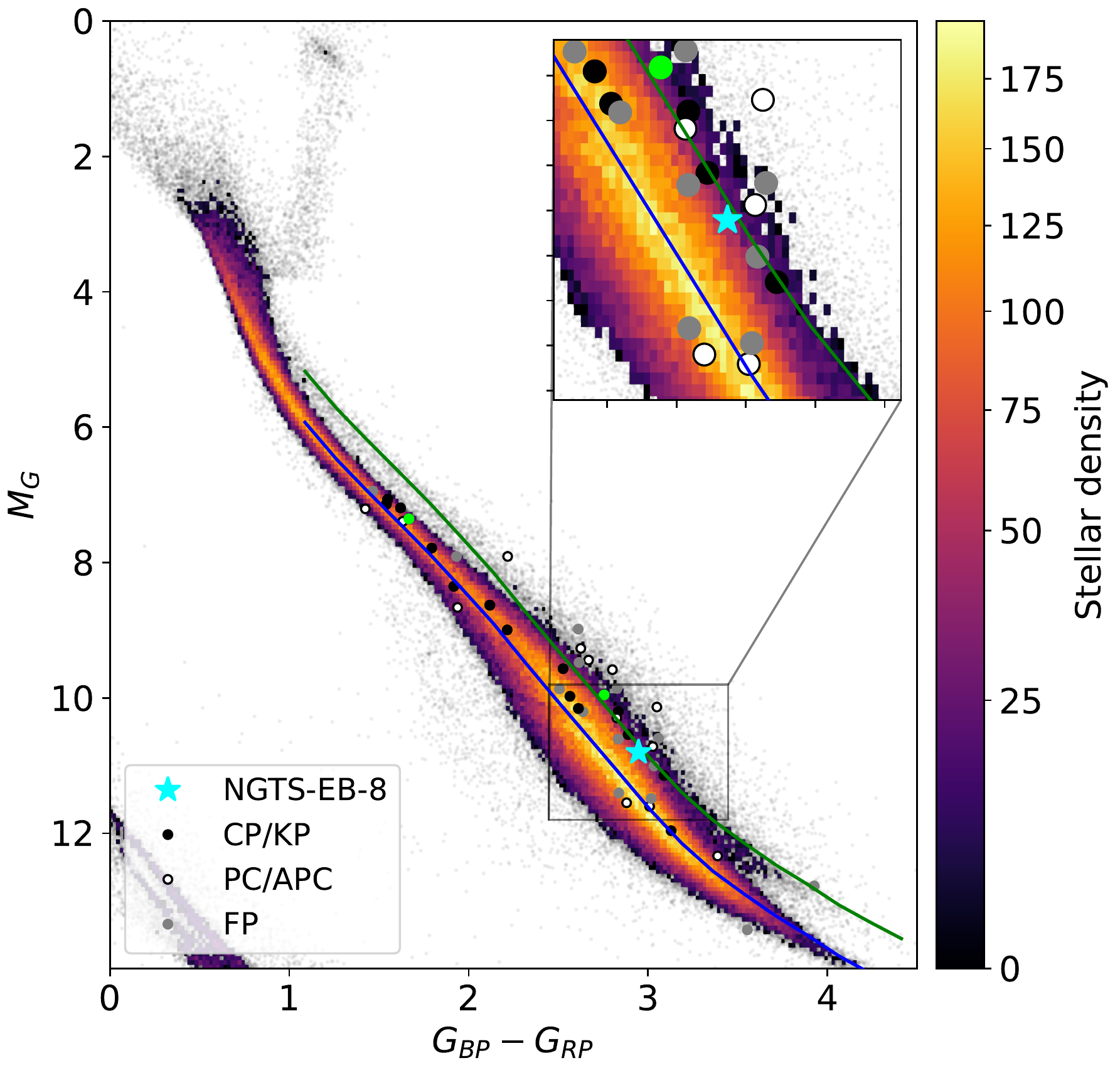}
    \caption{Gaia DR3 HR Diagrams for all stars within 100\,pc with the median fiducial tracking the main-sequence in blue and the same fiducial shifted by -0.753\,mag in green, corresponding to the position of unresolved binaries composed of two identical stars. \nigelsysname\ is shown as a cyan star and the inset shows the region zoomed in around \nigelsysname. Low-mass host stars of confirmed exoplanets, TOIs and candidates identified by \citetalias{Bryant2023OccurrenceGiantsMDwarfs} are overplotted, colored by TFOPWG/\citetalias{Bryant2023OccurrenceGiantsMDwarfs} disposition. Confirmed/Known Planets are shown as black filled circles, (Ambiguous) Planetary Candidates are shown as open black circles and False Positives are shown as grey circles. Binary systems hosting confirmed planets, Kepler-16 and TOI-762, are shown as lime green filled circles.}
    \label{fig:gaiahr}
\end{figure}

Given the importance of ruling out eclipsing binaries masquerading as planets in the context of measuring planet occurrence rates \citep[e.g.][]{Fressin2013KeplerFPandOccurrRate,Desert2015KeplerFP,Santerne2016SOPHIEKeplerGiantProperties,Triaud2017EBLM,Collins2018KELT_FPCat,TardugnoPoleo2024NotPlaNET}, we compare the properties of \nigelsysname\ to a population of confirmed and candidate planets.
We plot the position of \nigelsysname\ on the local ($d<100$\,pc) Gaia DR3 Hertzsprung–Russell (HR) diagram in Figure~\ref{fig:gaiahr}. Following the approach of \citet{GaiaCollaboration2018DR2HRD}, we overplot fiducial lines showing the main-sequence of the population (blue solid line) and the fiducial shifted by -0.753\,mag (green solid line) that corresponds to two identical stars in an unresolved binary system that will display the same color but twice the luminosity of the equivalent single star. 
We also show the positions of low-mass stars that: host confirmed exoplanets\footnote{\url{https://exoplanetarchive.ipac.caltech.edu/}} \citep{Akeson2013NASAExoArchive,Christiansen20225000Exoplanets}; have been flagged as \tess\ Objects of Interest \citep[TOIs;][]{Guerrero2021TOIRelease}; or were identified by \citetalias{Bryant2023OccurrenceGiantsMDwarfs}. We use the same effective temperature, stellar radius and distance criteria for low-mass stars as \citetalias{Bryant2023OccurrenceGiantsMDwarfs} when selecting the exoplanet hosts and TOIs, i.e. \teff$<4500$\,K, $R_*<0.75$\,\rsun\ and $d<100$\,pc. We maintain the $d<100$\,pc cut from \citetalias{Bryant2023OccurrenceGiantsMDwarfs} as the Gaia HR diagram and fiducial lines presented by \citet{GaiaCollaboration2018DR2HRD} that we re-create in Figure~\ref{fig:gaiahr} also uses this distance criterion. We impose a lower bound on the measured planetary/secondary radii of $R_2>0.6$\,\rjup\ to focus on the giant planet regime across these data. We also place an upper bound on the planetary mass for confirmed exoplanets of $M_{\text{pl}}>13.7$\,\mjup, the typically adopted value for the deuterium-burning limit separating planets and brown dwarfs \citep[e.g.][]{Spiegel2011DeutLimit}.
Finally, we select hosts with confirmed exoplanets that have a planetary radius measured to better than 20\% precision.
The TOIs are colored by the \tess\ Follow-up Working Group \citep[TFOPWG;][]{Collins2018TFOP} disposition (CP/KP=Confirmed/Known Planet, (A)PC=Ambiguous Planet Candidate, FP=False Positive) or we adopt the disposition provided in \citetalias{Bryant2023OccurrenceGiantsMDwarfs} for remaining targets.

We see that \nigelsysname\ sits close to the green fiducial line, in agreement with the assessment of the system as an unresolved binary. However, we note that a number of stars that host confirmed/known planets (black filled circles) also lie close to the green fiducial line despite Kepler-16 \citep{Doyle2011Kepler16bFirstCircumbinary} and TOI-762\,A \citep{Hartman2024TOI762Ab} (highlighted in lime green) being the only known binary systems hosting planets in the sample. TOI-762\,A that lies close to the green fiducial line is a resolved binary system, while Kepler-16 that lies nearer to the blue main-sequence fiducial is an unresolved binary system, contrary to expectation. We highlight however that neither system consists of two identical stars, with mass ratios smaller than 0.55 in both cases.
Systems hosting hot Jupiters have been shown to exhibit a higher likelihood having wide stellar companions relative to field stars \citep{Knutson2014HJFriendsI,Ngo2015HJFriendsII,Piskorz2015HJFriendsIII,Ngo2016HJFriendsIV}. In addition, the planet–metallicity correlation \citep{Gonzalez1997PMC,Santos2004PMC,Mortier2013PMC,Adibekyan2019PMC}, together with the elevated positions of metal-rich stars on the HR diagram \citep[e.g.][]{GaiaCollaboration2018DR2HRD}, further complicates the ability to identify planet candidates and binaries from their position on the HR diagram.
Similarly, we note that the re-normalized unit weight error (ruwe) reported in Gaia DR3 for \nigelsysname\ (\nigelgid) is $\text{ruwe}=1.0357833$. This is indicative of a good astrometric solution as it is close to 1. Indeed it is below the typically reported thresholds for potential binarity of 1.4 \citep{Lindegren2018ruwe} and 1.25 \citep{Penoyre2022ruwe}.

\begin{figure}
    \centering
    \includegraphics[width=0.9\columnwidth]{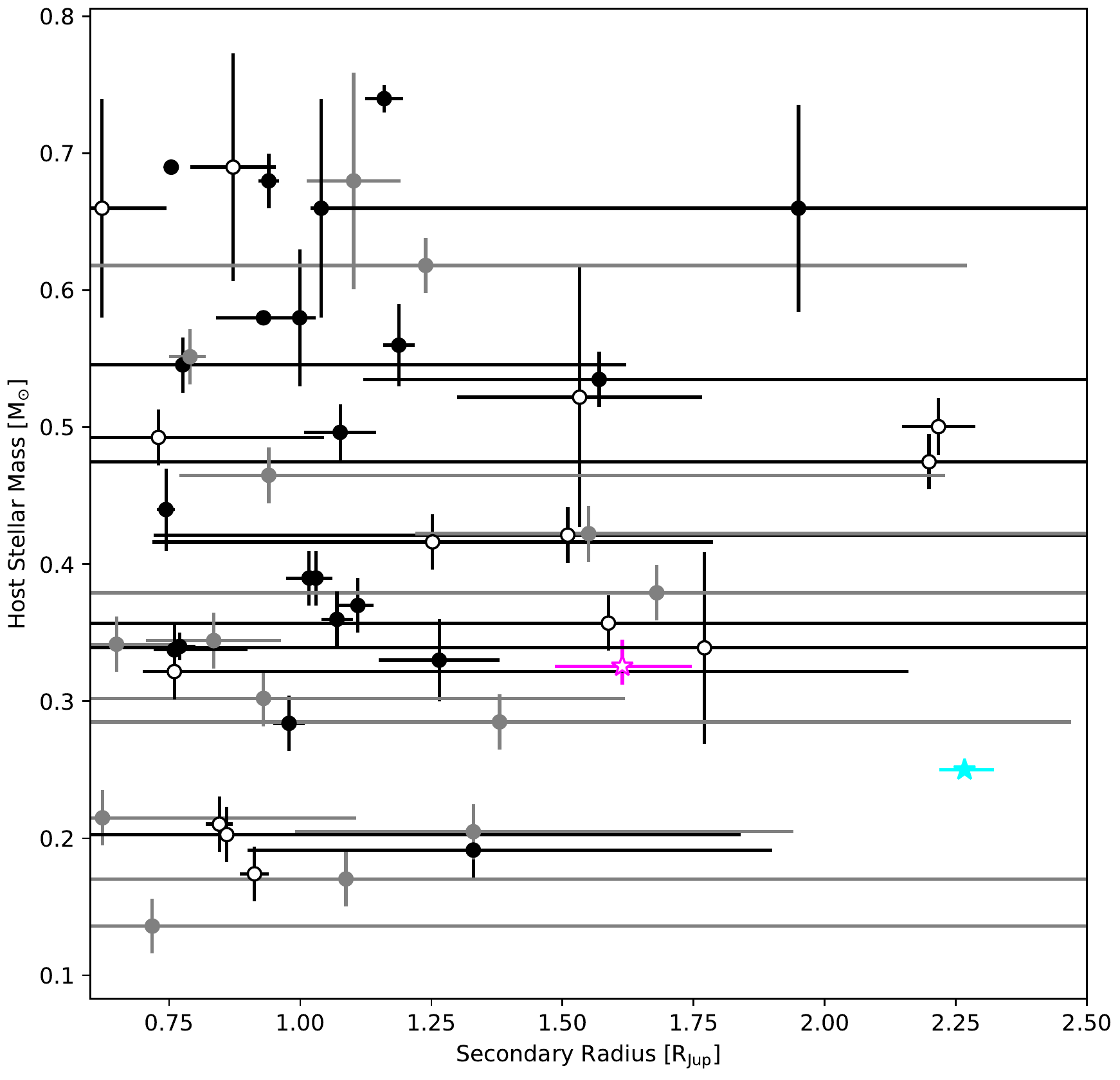}
    \caption{Secondary radius vs Host Stellar Mass for low-mass stars hosting confirmed exoplanets, TOIs and candidates identified in \citetalias{Bryant2023OccurrenceGiantsMDwarfs}. Coloured by disposition as described for Figure~\ref{fig:gaiahr}. The previously assumed position of \nigelsysname\ when it was described as a planet candidate in \citet{OBrien2024PHNGTSPaper1} is plotted as an unfilled magenta star and the position of \nigelsysname\ as determined  by the High $k$-range model from this work is shown as a filled cyan star.}
    \label{fig:mass_plrad}
\end{figure}

We also plot the secondary radius against host stellar mass for the same population of low-mass stars hosting exoplanets, TOIs and \citetalias{Bryant2023OccurrenceGiantsMDwarfs} candidates in Figure~\ref{fig:mass_plrad}. We plot the position of \nigelsysname\ using the stellar mass and secondary radius quoted in \citet{OBrien2024PHNGTSPaper1} as an unfilled cyan star, to give an indication of the perceived location of this system when it was deemed a planetary candidate. The revised position determined by the High $k$-range model from this work is also plotted.
While occupying an underpopulated region of the parameter space, the position of \nigelsysname\ as determined in \citet{OBrien2024PHNGTSPaper1} is not anomalous compared to the population of confirmed planets.
The confirmation or ruling out of planetary candidates in the very low-mass regime is crucial to further constraining the occurrence rates in this region of the parameter space. As demonstrated by the data obtained for \nigelsysname\ and its position relative to the population of confirmed planets, false positives and planet candidates, it is clear that high-resolution reconnaissance spectra are necessary for the definitive identification of double-lined binaries in systems such as this.

\section{Conclusions}\label{sec:conclusions}
We present the identification and characterization of the \nigelsysname\ system. This double-lined, eclipsing binary system is composed of two fully convective M-dwarfs orbiting each other every $4.19$\,days. This system was initially identified as a giant planet candidate orbiting a low-mass star independently by \citetalias{Bryant2023OccurrenceGiantsMDwarfs} and by citizen scientists through the Planet Hunters NGTS project \citep{OBrien2024PHNGTSPaper1}.
Photometric observations and speckle imaging were unable to definitively identify the true nature of this system. High-resolution spectroscopic observations using the GHOST instrument mounted on the 8.1\,m Gemini-South telescope and subsequent analysis of NIRPS radial velocity measurements were crucial in identifying this system as a double-lined binary.
Sufficient signal-to-noise to identify this system as a double-lined binary was achieved with a modest time allocation (700\,second) for this relatively faint ($V\mysim16$\,mag) target, thus the efficient vetting of these faint targets with short reconnaissance observations is highly feasible.
The ruling out of this system, and other eclipsing binaries that display similar transit signals, as false positives are important steps in further constraining the occurrence rates of giant planets orbiting low-mass stars.
We detect Balmer line emission from at least one component of \nigelsysname, indicating some level of stellar activity, however we do not detect significant flaring and suggest further photometric monitoring to assess the flare activity of the components in this system.
The detection of the eclipses in ATLAS data raises future prospects for the detection of deep transiting giant planet candidates and eclipsing binaries in all-sky surveys such as ATLAS, ZTF, Pan-STARRS and, in the near future, LSST.
\nigelsysname\ can provide a valuable benchmark system for testing stellar evolution models and probing the M-dwarf radius inflation problem. However, further observations to more precisely determine the stellar parameters of the components are necessary to place this system more precisely in the context of the low-mass star population and determine whether or not the parameters of \nigelsysname\ agree with theoretical models.

\begin{acknowledgments}

The data presented in this paper are the result of the efforts of the \phngts\ volunteers, without whom this work would not have been possible. Their contributions are individually acknowledged at: \url{https://www.zooniverse.org/projects/mschwamb/planet-hunters-ngts/about/results}. The volunteers who classified this system at any stage of the \phngts\ project are included as co-authors to give them credit for their crucial role in this discovery. We also thank our \phngts\ Talk moderators Arttu Sainio and See Min Lim for their time and efforts helping the \phngts\ volunteer community.

We thank the anonymous reviewer for their thoughtful review of our manuscript that significantly helped refine our analysis.

\textbf{Data access:} All \phngts\ data supporting this study is provided as supplementary information accompanying the \phngts\ overview paper \citep{OBrien2024PHNGTSPaper1}. Some of the data presented in this paper were obtained from the Mikulski Archive for Space Telescopes (MAST) at the Space Telescope Science Institute. The specific observations analyzed can be accessed via the following references: QLP: \citet{Huang2020QLPdoi}; TIC: \citet{STScI2018TIC}. STScI is operated by the Association of Universities for Research in Astronomy, Inc., under NASA contract NAS5–26555. Support to MAST for these data is provided by the NASA Office of Space Science via grant NAG5–7584 and by other grants and contracts. Data from the NASA Exoplanet Archive Planetary Systems Table can be accessed via: \dataset[https://doi.org/10.26133/NEA12]{https://doi.org/10.26133/NEA12} \citep{nasaexoarchiveDOI}. Data from the Exoplanet Follow-up Observing Program (ExoFOP) Web Service can be accessed via: \dataset[https://doi.org/10.26134/ExoFOP5]{https://doi.org/10.26134/ExoFOP5} \citep{exofopDOI}.
Gemini/GHOST data are available through the \href{https://archive.gemini.edu/searchform}{Gemini Observatory Archive}.
ESO/NIRPS and NGTS data are available through the \href{https://archive.eso.org/cms.html}{ESO Science Archive Facility}.
All photometric data are provided in the supplementary material.

S.M.O. is supported by a UK Science and Technology Facilities Council (STFC) Studentship (ST/W507751/1). 
C.A.W. would like to acknowledge support from STFC (grant number ST/X00094X/1).
The contributions at the Mullard Space Science Laboratory by E.M.B. have been supported by STFC through the consolidated grant ST/W001136/1.
The contributions at the University of Warwick by S.G., P.J.W. and R.G.W. have been supported by STFC through consolidated grants ST/P000495/1, ST/T000406/1 and ST/X001121/1.
JSJ gratefully acknowledges support by FONDECYT grant 1240738 and from the ANID BASAL project FB210003.
K.L.H. is supported by a UK Science and Technology Facilities Council (STFC) Studentship (ST/X508706/1). 
T.R. is supported by a UK Science and Technology Facilities Council (STFC) Studentship (ST/Y509504/1).
SS acknowledges Fondo Comit´e Mixto-ESO Chile ORP 025/2022.

We thank Matt Nicholl for a useful discussion on nested sampling.
We thank Harry\,J.\,Greatorex for a useful discussion on stellar flares.
We thank Mitchell\,E.\,Young for helpful discussions on PHOENIX models.
We thank Storm Colloms for their help with model comparison using Bayesian evidence.

Based on data collected under the NGTS project at the ESO La Silla Paranal Observatory. The NGTS facility is operated by the consortium institutes with support from the UK Science and Technology Facilities Council (STFC) under projects ST/M001962/1, ST/S002642/1 and ST/W003163/1.

This publication uses data generated via the \href{https://www.zooniverse.org/}{Zooniverse.org} platform, development of which is funded by generous support, including from the National Science Foundation, NASA, the Institute of Museum and Library Services, UKRI, a Global Impact Award from Google, and the Alfred P. Sloan Foundation. The code base for the Zooniverse Project Builder Platform is available under an open-source license at \url{https://github.com/zooniverse/Panoptes} and \url{https://github.com/zooniverse/Panoptes-Front-End}.

This paper includes data collected by the TESS mission, which are publicly available from the Mikulski Archive for Space Telescopes (MAST). Funding for the TESS mission is provided by the NASA's Science Mission Directorate.

This paper uses observations made at the South African Astronomical Observatory (SAAO).

Based on observations obtained at the international Gemini Observatory, a program of NSF NOIRLab, which is managed by the Association of Universities for Research in Astronomy (AURA) under a cooperative agreement with the U.S. National Science Foundation on behalf of the Gemini Observatory partnership: the U.S. National Science Foundation (United States), National Research Council (Canada), Agencia Nacional de Investigaci\'{o}n y Desarrollo (Chile), Ministerio de Ciencia, Tecnolog\'{i}a e Innovaci\'{o}n (Argentina), Minist\'{e}rio da Ci\^{e}ncia, Tecnologia, Inova\c{c}\~{o}es e Comunica\c{c}\~{o}es (Brazil), and Korea Astronomy and Space Science Institute (Republic of Korea). Some of the observations in the paper made use of the High-Resolution Imaging instrument Zorro. Zorro was funded by the NASA Exoplanet Exploration Program and built at the NASA Ames Research Center by Steve B. Howell, Nic Scott, Elliott P. Horch, and Emmett Quigley. Zorro was mounted on the Gemini South telescope of the international Gemini Observatory. Program IDs: GS-2022A-FT-209; GS-24A-FT-107

Based on observations collected at the European Southern Observatory under ESO programme 111.254E.

This work has made use of data from the Asteroid Terrestrial-impact Last Alert System (ATLAS) project. The Asteroid Terrestrial-impact Last Alert System (ATLAS) project is primarily funded to search for near earth asteroids through NASA grants NN12AR55G, 80NSSC18K0284, and 80NSSC18K1575; byproducts of the NEO search include images and catalogs from the survey area. This work was partially funded by Kepler/K2 grant J1944/80NSSC19K0112 and HST GO-15889, and STFC grants ST/T000198/1 and ST/S006109/1. The ATLAS science products have been made possible through the contributions of the University of Hawaii Institute for Astronomy, the Queen’s University Belfast, the Space Telescope Science Institute, the South African Astronomical Observatory, and The Millennium Institute of Astrophysics (MAS), Chile.

This work has made use of data from the European Space Agency (ESA) mission {\it Gaia} (\url{https://www.cosmos.esa.int/gaia}), processed by the {\it Gaia} Data Processing and Analysis Consortium (DPAC, \url{https://www.cosmos.esa.int/web/gaia/dpac/consortium}). Funding for the DPAC has been provided by national institutions, in particular the institutions participating in the {\it Gaia} Multilateral Agreement.

This publication makes use of data products from the Two Micron All Sky Survey, which is a joint project of the University of Massachusetts and the Infrared Processing and Analysis Center/California Institute of Technology, funded by the National Aeronautics and Space Administration and the National Science Foundation.

The national facility capability for SkyMapper has been funded through ARC LIEF grant LE130100104 from the Australian Research Council, awarded to the University of Sydney, the Australian National University, Swinburne University of Technology, the University of Queensland, the University of Western Australia, the University of Melbourne, Curtin University of Technology, Monash University and the Australian Astronomical Observatory. SkyMapper is owned and operated by The Australian National University's Research School of Astronomy and Astrophysics. The survey data were processed and provided by the SkyMapper Team at ANU. The SkyMapper node of the All-Sky Virtual Observatory (ASVO) is hosted at the National Computational Infrastructure (NCI). Development and support of the SkyMapper node of the ASVO has been funded in part by Astronomy Australia Limited (AAL) and the Australian Government through the Commonwealth's Education Investment Fund (EIF) and National Collaborative Research Infrastructure Strategy (NCRIS), particularly the National eResearch Collaboration Tools and Resources (NeCTAR) and the Australian National Data Service Projects (ANDS).

This publication makes use of data products from the Wide-field Infrared Survey Explorer, which is a joint project of the University of California, Los Angeles, and the Jet Propulsion Laboratory/California Institute of Technology, and NEOWISE, which is a project of the Jet Propulsion Laboratory/California Institute of Technology. WISE and NEOWISE are funded by the National Aeronautics and Space Administration.

This research made use of Lightkurve, a Python package for Kepler and TESS data analysis \citep{Lightkurve2018lightkurve}.

This research has made use of NASA’s Astrophysics Data System.

This research has made use of the Spanish Virtual Observatory (https://svo.cab.inta-csic.es) project funded by MCIN/AEI/10.13039/501100011033/ through grant PID2020-112949GB-I00.


\end{acknowledgments}

\facilities{NGTS, \tess, SAAO:1.0m (SHOC), ATLAS, Gemini:South (Zorro), Gemini:South (GHOST), ESO:3.6m (NIRPS)}


\software{
astropy \citep{Astropy2013,Astropy2018v2,Astropy2022v5},
matplotlib \citep{Hunter2007matplotlib},
Jupyter Notebook \citep{Kluyver2016Jupyter},
NumPy \citep{Walt2011Numpy,Harris2020Numpy},
pandas \citep{pandas2023},
scipy \citep{Virtanen2020Scipy},
\href{https://github.com/zooniverse/Panoptes}{panoptes},
\href{https://github.com/zooniverse/panoptes-python-client}{panoptes-python-client},
lightkurve \citep{Lightkurve2018lightkurve},
eleanor \citep{Feinstein2019eleanor},
tess-point \citep{Burke2020tesspoint},
TESSCut \citep{Brasseur2019TESSCut},
NASA GSFC-eleanor lite \citep{Powell2022NASAGSFCeleanor},
allesfitter \citep{Guenther2019allesfittercode,Guenther2021allesfitterpaper},
ellc \citep{Maxted2016ellc},
emcee \citep{ForemanMackey2013emcee},
dynesty \citep{Speagle2020dynesty},
celerite \citep{ForemanMackey2017celerite}
}

\appendix

\section{Spectral fitting}\label{sec:spec_fit}
\begin{figure*}
    \centering
    \includegraphics[width=0.95\textwidth, height =0.9\textheight,keepaspectratio]{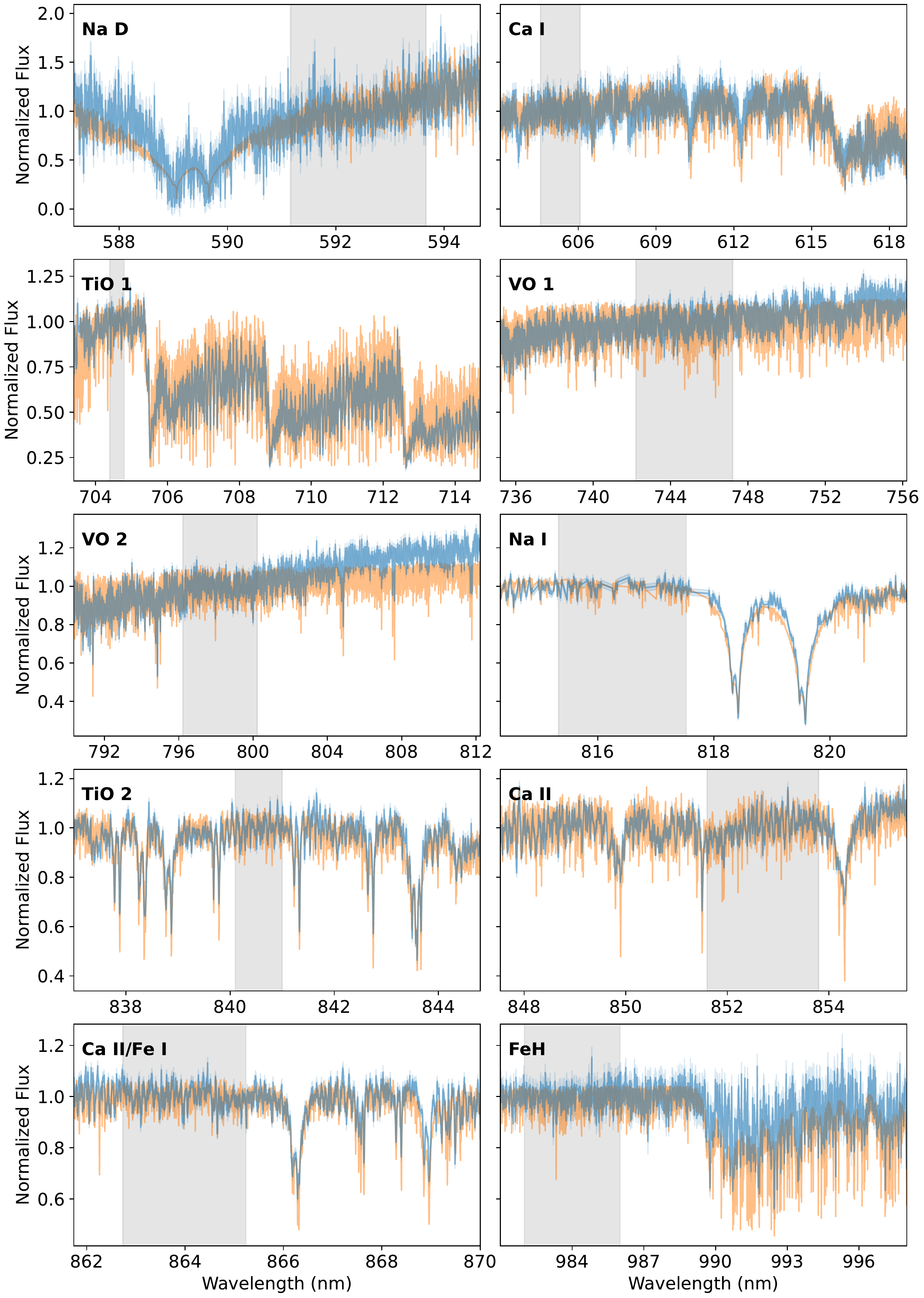}
    \caption{Spectral fits for each region defined in Table~\ref{tab:spec_regions}. The first GHOST spectrum of \nigelsysname\ is shown in blue. The best fitting SB2 template is shown in orange. The grey regions in each panel show the normalization range.}\label{fig:spec_fit}
\end{figure*}

Figure~\ref{fig:spec_fit} shows the best fitting synthetic SB2 template for the High $k$-range model. The parameters of each component are: $T_{\rm eff,A} = 3500$\,K, $\log g_A = 5.50$, $\mbox{[Fe/H]}_A = 0.0$\,dex, $T_{\rm eff,B} = 3200$\,K, $\log g_B = 5.50$, $\mbox{[Fe/H]}_B = 0.0$\,dex.

\section{Population of M+M Double-Lined Eclipsing Binaries}\label{sec:mm_table}
In Table~\ref{tab:mm_table} we provide the parameters and references for the M+M Double-Lined Eclipsing Binaries that we drew from literature. A machine-readable version of this table will be available in the journal version of this article.
\begin{deluxetable}{ccccccc}
\tabletypesize{\footnotesize}
\tablecaption{Parameters for double-lined eclipsing binaries composed of two M-dwarfs. \label{tab:mm_table}}
\tablehead{\colhead{Name} & \colhead{Period (days)} & \colhead{$M_A$ (\msun)} & \colhead{$M_B$ (\msun)} & \colhead{$R_A$ (\rsun)} & \colhead{$R_B$ (\rsun)} & \colhead{References} \\}
\startdata
Castor C & 0.8143 & $0.5992^{+0.0047}_{-0.0047}$ & $0.5992^{+0.0047}_{-0.0047}$ & $0.6191^{+0.0057}_{-0.0057}$ & $0.6191^{+0.0057}_{-0.0057}$ & (1),(2),(3),(4)\\
    WTS-17e-3-02003 & 1.225 & $0.597^{+0.02}_{-0.02}$ & $0.51^{+0.016}_{-0.016}$ & $0.611^{+0.095}_{-0.095}$ & $0.54^{+0.11}_{-0.11}$ & (5)\\
    MG1-506664 & 1.548 & $0.584^{+0.002}_{-0.002}$ & $0.544^{+0.002}_{-0.002}$ & $0.56^{+0.005}_{-0.005}$ & $0.513^{+0.008}_{-0.008}$ & (6),(7)\\
    NGTS J0930-18 & 1.333 & $0.5803^{+0.0092}_{-0.0063}$ & $0.0818^{+0.004}_{-0.0015}$ & $0.584^{+0.0094}_{-0.01}$ & $0.1059^{+0.0023}_{-0.0021}$ & (8)\\
    WTS-19G-4-02069B & 2.44 & $0.53^{+0.02}_{-0.02}$ & $0.143^{+0.006}_{-0.006}$ & $0.51^{+0.01}_{-0.01}$ & $0.174^{+0.006}_{-0.006}$ & (9)\\
    MG1-78457 & 1.586 & $0.527^{+0.002}_{-0.002}$ & $0.491^{+0.002}_{-0.002}$ & $0.505^{+0.011}_{-0.011}$ & $0.471^{+0.012}_{-0.012}$ & (6),(7)\\
    WTS-17h-4-01429 & 1.445 & $0.503^{+0.016}_{-0.016}$ & $0.409^{+0.013}_{-0.013}$ & $0.514^{+0.006}_{-0.006}$ & $0.421^{+0.006}_{-0.006}$ & (5)\\
    MG1-646680 & 1.638 & $0.499^{+0.002}_{-0.002}$ & $0.443^{+0.002}_{-0.002}$ & $0.457^{+0.007}_{-0.007}$ & $0.427^{+0.006}_{-0.006}$ & (6),(7)\\
    THOR 42 & 0.859 & $0.497^{+0.005}_{-0.005}$ & $0.205^{+0.002}_{-0.002}$ & $0.659^{+0.003}_{-0.003}$ & $0.424^{+0.002}_{-0.002}$ & (6),(10)\\
    MG1-2056316 & 1.722 & $0.469^{+0.002}_{-0.002}$ & $0.382^{+0.002}_{-0.002}$ & $0.441^{+0.003}_{-0.003}$ & $0.374^{+0.003}_{-0.003}$ & (6),(7)\\
    HAT-TR-318-007 & 3.343 & $0.448^{+0.001}_{-0.001}$ & $0.2721^{+0.0042}_{-0.0042}$ & $0.4548^{+0.0036}_{-0.0036}$ & $0.2913^{+0.0024}_{-0.0024}$ & (6),(11)\\
    CU Cnc & 2.771 & $0.4358^{+0.0008}_{-0.0008}$ & $0.3998^{+0.0014}_{-0.0014}$ & $0.4317^{+0.0052}_{-0.0052}$ & $0.3908^{+0.0094}_{-0.0094}$ & (6),(12),(13),(14),(15),(16)\\
    NGTS J214358.5-380102 & 7.618 & $0.426^{+0.0056}_{-0.0049}$ & $0.455^{+0.0058}_{-0.0052}$ & $0.461^{+0.038}_{-0.025}$ & $0.411^{+0.027}_{-0.039}$ & (17)\\
    NGTS 0002-29 & 1.098 & $0.3978^{+0.0033}_{-0.0033}$ & $0.2245^{+0.0018}_{-0.0018}$ & $0.4037^{+0.0048}_{-0.0048}$ & $0.2759^{+0.0055}_{-0.0055}$ & (6),(18)\\
    EPIC 211972086 & 6.016 & $0.3953^{+0.002}_{-0.002}$ & $0.2098^{+0.0014}_{-0.0014}$ & $0.363^{+0.008}_{-0.008}$ & $0.272^{+0.012}_{-0.012}$ & (6),(19)\\
    LSPM J1112+7626 & 41.03 & $0.3946^{+0.0023}_{-0.0023}$ & $0.2745^{+0.0012}_{-0.0012}$ & $0.386^{+0.0054}_{-0.0054}$ & $0.2978^{+0.0048}_{-0.0048}$ & (6),(20)\\
    WTS-19c-3-08647 & 0.8675 & $0.393^{+0.019}_{-0.019}$ & $0.244^{+0.014}_{-0.014}$ & $0.494^{+0.069}_{-0.069}$ & $0.421^{+0.006}_{-0.006}$ & (5)\\
    GJ 3236 & 0.7713 & $0.376^{+0.017}_{-0.017}$ & $0.281^{+0.015}_{-0.015}$ & $0.3828^{+0.007}_{-0.007}$ & $0.2992^{+0.0075}_{-0.0075}$ & (1),(21),(22)\\
    LP 133-373 & 1.6 & $0.34^{+0.014}_{-0.014}$ & $0.34^{+0.014}_{-0.014}$ & $0.33^{+0.02}_{-0.02}$ & $0.33^{+0.02}_{-0.02}$ & (23)\\
    UScoCTIO 5 & 34.0 & $0.3393^{+0.002}_{-0.002}$ & $0.3263^{+0.002}_{-0.002}$ & $0.8733^{+0.0024}_{-0.0024}$ & $0.8376^{+0.0024}_{-0.0024}$ & (6),(24),(25)\\
    LP 661-13 & 4.704 & $0.308^{+0.00084}_{-0.00084}$ & $0.194^{+0.00034}_{-0.00034}$ & $0.3226^{+0.0033}_{-0.0033}$ & $0.2174^{+0.0023}_{-0.0023}$ & (26)\\
    SDSS-MEB-1 & 0.407 & $0.272^{+0.02}_{-0.02}$ & $0.24^{+0.022}_{-0.022}$ & $0.268^{+0.01}_{-0.01}$ & $0.248^{+0.009}_{-0.009}$ & (27)\\
    1RXS-J1547+4508 & 3.55 & $0.258^{+0.009}_{-0.009}$ & $0.259^{+0.008}_{-0.008}$ & $0.289^{+0.007}_{-0.007}$ & $0.289^{+0.007}_{-0.007}$ & (28),(29)\\
    LP 177-102 & 3.55 & $0.2576^{+0.0085}_{-0.0085}$ & $0.2585^{+0.008}_{-0.008}$ & $0.2895^{+0.0068}_{-0.0068}$ & $0.2895^{+0.0068}_{-0.0068}$ & (1),(30),(29),(31)\\
    HATS551-027 & 4.077 & $0.244^{+0.003}_{-0.003}$ & $0.179^{+0.002}_{-0.001}$ & $0.261^{+0.006}_{-0.009}$ & $0.218^{+0.007}_{-0.011}$ & (32)\\
    KOI-126 BC & 1.722 & $0.2352^{+0.0006}_{-0.0006}$ & $0.2073^{+0.0006}_{-0.0006}$ & $0.2545^{+0.0008}_{-0.0008}$ & $0.2315^{+0.0007}_{-0.0007}$ & (6),(33),(34)\\
    CM Dra & 1.268 & $0.225^{+0.0003}_{-0.0003}$ & $0.2101^{+0.0003}_{-0.0003}$ & $0.251^{+0.0002}_{-0.0002}$ & $0.2375^{+0.0002}_{-0.0002}$ & (6),(35),(36),(14),(37),(38),(39)\\
    TOI-450 & 10.71 & $0.1768^{+0.0004}_{-0.0004}$ & $0.1767^{+0.0003}_{-0.0003}$ & $0.345^{+0.006}_{-0.006}$ & $0.346^{+0.006}_{-0.006}$ & (6),(40)\\
    NGTS J052218.2-250710.4 & 1.748 & $0.1739^{+0.0013}_{-0.0013}$ & $0.1742^{+0.0013}_{-0.0013}$ & $0.2045^{+0.0048}_{-0.0048}$ & $0.2168^{+0.0048}_{-0.0048}$ & (6),(41)\\
    TMTSJ0803 & 0.2217 & $0.169^{+0.01}_{-0.01}$ & $0.162^{+0.016}_{-0.016}$ & $0.17^{+0.006}_{-0.006}$ & $0.156^{+0.006}_{-0.006}$ & (42)\\
    EPIC 203710387 & 2.809 & $0.1158^{+0.0031}_{-0.0031}$ & $0.1056^{+0.0027}_{-0.0027}$ & $0.4317^{+0.0055}_{-0.0055}$ & $0.4236^{+0.0056}_{-0.0056}$ & (6),(43),(44)\\
\enddata
\tablecomments{(1): \citep{2025AnA...693A.228C}, (2): \citep{1926ApJ....64..250J}, (3): \citep{2002AJ....123.3356G}, (4): \citep{2022ApJ...941....8T}, (5): \citep{2018MNRAS.476.5253C}, (6): \citep{2015ASPC..496..164S}, (7): \citep{2011ApJ...728...48K}, (8): \citep{2020MNRAS.498.3115A}, (9): \citep{2013MNRAS.431.3240N}, (10): \citep{2020MNRAS.491.4902M}, (11): \citep{2018AJ....155..114H}, (12): \citep{1999AnA...344..897D}, (13): \citep{1999AnA...341L..63D}, (14): \citep{2010AnARv..18...67T}, (15): \citep{2024AnA...684A.175H}, (16): \citep{2003AnA...398..239R}, (17): \citep{2020MNRAS.494.3950A}, (18): \citep{2021MNRAS.507.5991S}, (19): \citep{2017ApJ...845...72K}, (20): \citep{2011ApJ...742..123I}, (21): \citep{2009ApJ...701.1436I}, (22): \citep{2010ApJ...716.1522S}, (23): \citep{2007ApJ...661.1112V}, (24): \citep{2015ApJ...807....3K}, (25): \citep{2019ApJ...872..161D}, (26): \citep{2017ApJ...836..124D}, (27): \citep{2008ApJ...684..635B}, (28): \citep{2023Univ....9..498M}, (29): \citep{2011AJ....141..166H}, (30): \citep{2002AJ....124.2868M}, (31): \citep{2012MNRAS.426.1507B}, (32): \citep{2015MNRAS.451.2263Z}, (33): \citep{2011Sci...331..562C}, (34): \citep{2022ApJ...924...66Y}, (35): \citep{1977ApJ...218..444L}, (36): \citep{2009ApJ...691.1400M}, (37): \citep{2019MNRAS.488.4905S}, (38): \citep{2012ApJ...760L...9T}, (39): \citep{2024MNRAS.528..963M}, (40): \citep{2023AJ....165...46T}, (41): \citep{2018MNRAS.481.1897C}, (42): \citep{2024MNRAS.531.1765L}, (43): \citep{2015AnA...584A.128L}, (44): \citep{2016ApJ...816...21D}}
\end{deluxetable}

\bibliography{zbiblio}{}
\bibliographystyle{aasjournalv7}



\end{document}